# Five New and Three Improved Mutual Orbits of Transneptunian Binaries

W.M. Grundy[a], K.S. Noll[b], F. Nimmo[c], H.G. Roe[a], M.W. Buie[d],
S.B. Porter[a], S.D. Benecchi[e,f], D.C. Stephens[g], H.F. Levison[d], and J.A. Stansberry[h]

a. Lowell Observatory, 1400 W. Mars Hill Rd., Flagstaff AZ 86001.
b. Space Telescope Science Institute, 3700 San Martin Dr., Baltimore MD 21218.
c. Department of Earth and Planetary Sciences, University of California, Santa Cruz CA 95064.
d. Southwest Research Institute, 1050 Walnut St. #300, Boulder CO 80302.
e. Planetary Science Institute, 1700 E. Fort Lowell, Suite 106, Tucson AZ 85719.
f. Carnegie Institution of Washington, Department of Terrestrial Magnetism, 5241 Broad Branch Rd. NW, Washington DC 20015.
g. Dept. of Physics and Astronomy, Brigham Young University, N283 ESC Provo UT 84602.
h. Steward Observatory, University of Arizona, 933 N. Cherry Ave., Tucson AZ 85721.

**ABSTRACT**

We present three improved and five new mutual orbits of transneptunian binary systems (58534) Logos-Zoe, (66652) Borasisi-Pabu, (88611) Teharonhiawako-Sawiskera, (123509) 2000 WK$_{183}$, (149780) Altjira, 2001 QY$_{297}$, 2003 QW$_{111}$, and 2003 QY$_{90}$ based on Hubble Space Telescope and Keck 2 laser guide star adaptive optics observations. Combining the five new orbit solutions with 17 previously known orbits yields a sample of 22 mutual orbits for which the period *P*, semimajor axis *a*, and eccentricity *e* have been determined. These orbits have mutual periods ranging from 5 to over 800 days, semimajor axes ranging from 1,600 to 37,000 km, eccentricities ranging from 0 to 0.8, and system masses ranging from $2 \times 10^{17}$ to $2 \times 10^{22}$ kg. Based on the relative brightnesses of primaries and secondaries, most of these systems consist of near equal-sized pairs, although a few of the most massive systems are more lopsided. The observed distribution of orbital properties suggests that the most loosely-bound transneptunian binary systems are only found on dynamically cold heliocentric orbits. Of the 22 known binary mutual orbits, orientation ambiguities are now resolved for 9, of which 7 are prograde and 2 are retrograde, consistent with a random distribution of orbital orientations, but not with models predicting a strong preference for retrograde orbits. To the extent that other perturbations are not dominant, the binary systems undergo Kozai oscillations of their eccentricities and inclinations with periods of the order of tens of thousands to millions of years, some with strikingly high amplitudes.

# 1. Introduction

Transneptunian objects (TNOs) record valuable information about the chemical and physical conditions in the outer parts of the protoplanetary nebula where they formed. Since spacecraft access to study their compositions and interior structures is severely limited by their extreme distances, telescopic observations are the only way to study a large sample of TNOs. Their great distances and small sizes limit what can be done using even the most capable telescope facilities. Fortunately, the existence of numerous transneptunian binaries (TNBs) provides a way of learning about their bulk properties via remote observations (e.g., Noll et al. 2008a). They also enable comparisons between TNBs belonging to the various dynamical sub-classes (e.g., Elliot et al. 2005; Gladman et al. 2008). These include "Classical" objects on low inclination, low eccentricity orbits, "Scattered" objects occupying more excited orbits, and "Resonant" objects trapped in a variety of mean motion resonances with Neptune.

The sample of TNBs with known mutual orbits has expanded rapidly in recent years (see Table 1). Remote observation of their mutual orbital semimajor axes and periods gives their total system masses, along with many other properties that would be otherwise unobtainable. The orbits of a large ensemble of binary systems can be used to place additional constraints on possible formation mechanisms as well as subsequent dynamical history. This paper adds five more systems to that sample and improves the orbits of three others.

**Table 1.**
Heliocentric orbital characteristics of TNBs with known orbits

| TNB system number, designation, and name | Mean heliocentric orbital elements[a] | | | Dynamical class[b] |
|---|---|---|---|---|
| | $a_\odot$ (AU) | $e_\odot$ | $i_\odot$ (°) | |
| *Systems with new orbits* | | | | |
| (123509) 2000 WK$_{183}$ | 44.4 | 0.048 | 2.72 | Classical |
| (148780) 2001 UQ$_{18}$ Altjira | 44.3 | 0.059 | 5.47 | Classical |
| 2001 QY$_{297}$ | 43.9 | 0.074 | 0.96 | Classical |
| 2003 QW$_{111}$ | 43.7 | 0.109 | 1.27 | Resonant 7:4 |
| 2003 QY$_{90}$ | 42.8 | 0.057 | 2.21 | Classical |
| *Systems with improved orbits[c]* | | | | |
| (58534) 1997 CQ$_{29}$ Logos | 45.2 | 0.125 | 2.01 | Classical |
| (66652) 1999 RZ$_{253}$ Borasisi | 43.8 | 0.080 | 1.57 | Classical |
| (88611) 2001 QT$_{297}$ Teharonhiawako | 44.1 | 0.027 | 4.18 | Classical |
| *Systems with published orbits[c]* | | | | |
| (26308) 1998 SM$_{165}$ | 47.8 | 0.375 | 13.08 | Resonant 2:1 |
| (42355) 2002 CR$_{46}$ Typhon | 38.1 | 0.538 | 3.79 | Centaur |
| (65489) 2003 FX$_{128}$ Ceto | 105.4 | 0.831 | 21.44 | Centaur |
| (90482) 2004 DW Orcus | 39.5 | 0.254 | 21.19 | Resonant 3:2 |
| (120347) 2004 SB$_{60}$ Salacia | 42.1 | 0.104 | 25.57 | Extended scattered |
| (134860) 2000 OJ$_{67}$ | 42.9 | 0.013 | 1.32 | Classical |

| | | | | | |
|---|---|---|---|---|---|
| (136199) | 2003 UB$_{313}$ Eris | 67.9 | 0.446 | 43.22 | Extended scattered |
| | 1998 WW$_{31}$ | 44.7 | 0.085 | 8.34 | Classical |
| | 1999 OJ4 | 38.1 | 0.018 | 2.58 | Classical |
| | 2000 QL$_{251}$ | 47.8 | 0.208 | 5.80 | Resonant 2:1 |
| | 2001 QC$_{298}$ | 46.3 | 0.128 | 31.54 | Extended scattered |
| | 2001 XR$_{254}$ | 43.0 | 0.024 | 2.66 | Classical |
| | 2003 TJ$_{58}$ | 44.5 | 0.094 | 1.31 | Classical |
| | 2004 PB$_{108}$ | 45.1 | 0.107 | 19.19 | Extended scattered |

Table notes:

[a.] Averaged over a 10 Myr integration, with $i_\odot$ relative to the invariable plane as described by Elliot et al. (2005).

[b.] Classifications are according to the current Deep Ecliptic Survey system (DES; see links from `http://www.boulder.swri.edu/~buie/kbo`; the original DES classification scheme was described by Elliot et al. 2005 and a manuscript detailing minor subsequent revisions is in preparation). The Gladman et al. (2008) system would classify these objects much the same, except for Salacia and 2001 QC$_{298}$ which are considered Classical in that system and Eris, which would be classed as detached.

[c.] Orbits for these systems have been reported by Veillet et al. (2002), Noll et al. (2004a,b), Brown and Schaller (2007), Grundy et al. (2007, 2008, 2009), Brown et al. (2010), and Stansberry et al. (2011).

## 2. New and Improved Orbits

Data used in this paper to determine or improve TNB orbits were acquired using the Hubble Space Telescope (HST) and the Keck II telescope on Mauna Kea. Relevant HST observations were obtained through programs 9060, 9386, 9585, 9746, 9991, 10508, 10514, 10800, and 11178, extending over Cycles 10 through 16. These nine programs, led by several different investigators, employed a variety of instruments, filters, and observing strategies. Details of astrometric data reduction procedures for various HST programs and instruments are described elsewhere (Stephens and Noll 2006; Grundy et al. 2008, 2009). In general, relative astrometry was obtained by fitting a pair of Tiny Tim (e.g., Krist and Hook 2004) point-spread functions (PSFs) to each image, then estimating astrometric uncertainties from the scatter of the separate measurements obtained over the course of each HST visit to a particular system. An uncertainty floor was imposed to avoid over-weighting visits which could happen to have had small measurement scatters by chance. We set this floor to 1 mas for WFPC2/PC data and 0.5 mas for ACS/HRC data. The various filters, cameras, and integration times used in the nine HST programs resulted in a very heterogeneous photometric data set. For filters near *V* band (*F475W*, *F555W*, and *F606W*) where color information was also available, we converted the observed fluxes to *V* magnitudes, as described in detail by Benecchi et al. (2009).

Additional observations were done at Keck II using the NIRC2 camera with laser guide star adaptive optics (e.g., Le Mignant et al. 2006). These observations required the presence of a nearby (< 30 arcsec) and much brighter (*R* < 18 mag) appulse star for tip-tilt corrections. Target motion with respect to the appulse star was compensated for by use of a new differential tracking mode implemented by A. Conrad at Keck Observatory. The observations were done in an *H* band filter (1.49 – 1.78 µm), using stacks of three consecutive one to two minute integrations followed by a dither, then three more consecutive integrations, and so on. The idea behind recording groups of three frames was to enable us to co-add to reach better sensitivity, while pre-serving the ability to discard any frames happening to have poor image quality due to variable seeing conditions (which turned out to be a rare occurrence). As with the HST data, astrometric reduction of each stack of 3 frames was done by means of PSF fitting. We experimented with azimuthally symmetric Gaussian and Lorentzian profiles, and for each visit, selected the profile leading to the lowest $\chi^2$ for the PSF-fit. Most often, this was the Gaussian profile. Its width was fitted simultaneously with the positions of the two components of the binary. We assumed a mean plate scale of 9.963 mas/pixel and an orientation offset of 0.13° (e.g., Ghez et al. 2008; Konopacky et al. 2010). No photometric standards were observed, and no effort was made to compute $H$ band magnitudes from these data, which were taken solely for astrometric purposes.

Table 2 lists the mean relative astrometric measurements and estimated 1-σ uncertainties for the eight systems whose new or improved orbits are presented in this paper. Data from previously published observations are included in the form used in our orbit fits. Observations available in the HST archive were re-reduced using our current pipeline, in order to be as consistent as possible, so the numbers in this table may not exactly agree with previously published numbers from the same data. We also include separate visual photometry for primary and secondary bodies, when available. Visual brightness differences between primaries and secondaries are mostly less than a magnitude, indicative of pairs of similar-sized bodies, but a few systems show magnitude differences greater than two.

**Table 2.**

Astrometric and photometric data used,
including new Hubble Space Telescope and Keck observations

| System and mean UT observation date and hour | Instrument[a] or source | $r$[b] | $\Delta$[b] | Phase angle[b] | $\Delta x$[c] | $\Delta y$[c] | $V_{primary}$ | $V_{secondary}$ |
|---|---|---|---|---|---|---|---|---|
| | | (AU) | | (°) | (arcsec) | | (magnitudes)[d] | |
| **(58534) Logos-Zoe** | | | | | | | | |
| 2001/11/17  6.9600 | WFPC2(WF3) | 41.564 | 41.861 | 1.30 | +0.055(16) | +0.192(29) | 23.09(10) | 23.48(15) |
| 2002/06/18  8.0322 | WFPC2 | 41.618 | 41.880 | 1.35 | −0.1464(18) | +0.3001(15) | - | - |
| 2002/06/30  3.6822 | WFPC2 | 41.621 | 42.068 | 1.25 | −0.1277(17) | +0.3109(10) | - | - |
| 2002/07/12  4.0405 | WFPC2 | 41.624 | 42.242 | 1.10 | −0.1040(10) | +0.3181(38) | - | - |
| 2003/05/04  4.7256 | NICMOS | 41.700 | 41.197 | 1.21 | −0.1320(10) | +0.3182(76) | - | - |
| 2004/02/16 17.7899 | ACS | 41.775 | 40.835 | 0.42 | −0.14768(53) | +0.28121(70) | - | - |
| 2004/05/01 12.7672 | ACS | 41.794 | 41.242 | 1.16 | −0.0204(52) | +0.3050(13) | - | - |
| 2004/06/23  1.3136 | ACS | 41.808 | 42.107 | 1.33 | +0.0692(17) | +0.01995(95) | - | - |
| 2007/12/17  7.1353 | WFPC2 | 42.141 | 42.072 | 1.34 | −0.0391(10) | −0.1613(27) | 24.143(38) | 24.595(42) |
| **(66652) Borasisi-Pabu** | | | | | | | | |
| 2003/04/23  3.4388 | NICMOS | 41.058 | 41.549 | 1.22 | +0.1747(18) | −0.0854(22) | - | - |
| 2003/08/20 14.7750 | ACS | 41.079 | 40.070 | 0.11 | −0.02455(73) | +0.07100(67) | 22.76(05) | 23.08(07) |
| 2003/09/15  6.0268 | ACS | 41.084 | 40.142 | 0.50 | +0.22368(52) | −0.04275(71) | 22.69(04) | 23.10(06) |
| 2003/11/17 12.0937 | ACS | 41.095 | 40.981 | 1.37 | +0.06458(50) | +0.07964(68) | 22.84(07) | 23.37(08) |
| 2003/11/29  7.3106 | ACS | 41.097 | 41.188 | 1.37 | −0.0175(11) | −0.07675(75) | 22.77(07) | 23.19(07) |
| 2007/07/17 16.1475 | WFPC2 | 41.345 | 40.600 | 0.97 | +0.1501(13) | +0.0662(10) | 22.990(30) | 23.292(41) |
| 2008/05/01 19.0186 | WFPC2 | 41.401 | 41.861 | 1.23 | −0.0797(12) | +0.0093(22) | 22.940(21) | 23.470(30) |
| **(88611) Teharonhiawako-Sawiskera** | | | | | | | | |
| 2001/10/11  0.9528 | Lit:O'03 | 44.955 | 44.370 | 1.04 | +0.5390(51) | −0.2770(52) | - | - |
| 2001/10/12  1.8730 | Lit:O'03 | 44.955 | 44.385 | 1.05 | +0.5460(81) | −0.2675(84) | - | - |
| 2001/11/01  0.7697 | Lit:O'03 | 44.956 | 44.699 | 1.23 | +0.624(21) | −0.214(24) | - | - |
| 2001/11/02  0.4299 | Lit:O'03 | 44.956 | 44.716 | 1.23 | +0.644(21) | −0.184(26) | - | - |
| 2001/11/03  0.3249 | Lit:O'03 | 44.956 | 44.733 | 1.24 | +0.642(40) | −0.193(39) | - | - |
| 2001/11/04  0.9046 | Lit:O'03 | 44.956 | 44.750 | 1.24 | +0.645(21) | −0.138(35) | - | - |
| 2002/07/13  6.7387 | Lit:O'03 | 44.966 | 44.130 | 0.75 | −0.314(23) | +0.692(29) | - | - |
| 2002/07/18  6.9538 | Lit:O'03 | 44.966 | 44.085 | 0.65 | −0.344(68) | +0.700(55) | - | - |
| 2002/08/07  4.5629 | Lit:O'03 | 44.967 | 43.970 | 0.24 | −0.43(13) | +0.81(13) | - | - |
| 2002/09/08  5.7632 | Lit:O'03 | 44.968 | 44.024 | 0.45 | −0.658(91) | +0.658(91) | - | - |
| 2003/10/23  1.7567 | Lit:K'05 | 44.985 | 44.536 | 1.14 | −0.012(60) | −0.527(50) | - | - |
| 2004/05/25  8.7890 | Lit:K'05 | 44.994 | 44.878 | 1.28 | +0.4350(70) | +0.4560(70) | - | - |
| 2004/09/13  3.2531 | Lit:K'05 | 44.999 | 44.074 | 0.51 | −0.1330(70) | +0.6990(60) | - | - |
| 2005/07/11  5.8782 | Lit:K'05 | 45.012 | 44.234 | 0.84 | −1.0020(80) | −0.0440(80) | - | - |
| 2009/12/12  5.2070 | Keck/NIRC2 | 45.081 | 45.368 | 1.19 | −1.0257(77) | +0.1098(53) | - | - |
| 2010/08/03 10.1942 | Keck/NIRC2 | 45.088 | 44.161 | 0.53 | −0.0032(30) | −0.5015(30) | - | - |

| System and mean UT observation date and hour | Instrument[a] or source | $r$[b] (AU) | $\Delta$[b] (AU) | Phase angle[b] (°) | $\Delta x$[c] (arcsec) | $\Delta y$[c] (arcsec) | $V_{primary}$ (magnitudes)[d] | $V_{secondary}$ (magnitudes)[d] |
|---|---|---|---|---|---|---|---|---|
| **(123509) 2000 WK$_{183}$** | | | | | | | | |
| 2005/11/24  9.1364 | ACS | 42.971 | 41.996 | 0.21 | +0.0541(22) | −0.05777(50) | - | - |
| 2007/08/08 16.7433 | WFPC2 | 42.918 | 43.394 | 1.19 | +0.0739(13) | −0.0055(28) | 23.873(11) | 23.960(25) |
| 2007/08/13 17.5017 | WFPC2 | 42.918 | 43.317 | 1.24 | +0.04456(96) | +0.0620(20) | 23.840(48) | 24.158(78) |
| 2007/09/12 21.7158 | WFPC2 | 42.915 | 42.814 | 1.34 | +0.0487(15) | +0.0562(43) | 23.945(78) | 23.943(67) |
| 2007/10/16 20.1289 | WFPC2 | 42.913 | 42.283 | 1.04 | +0.0087(30) | +0.0760(11) | 23.893(28) | 23.939(22) |
| 2007/11/27 21.5019 | WFPC2 | 42.909 | 41.934 | 0.20 | −0.0639(21) | −0.0387(44) | 23.619(33) | 24.065(35) |
| 2008/08/22 14.1519 | WFPC2 | 42.888 | 43.156 | 1.30 | −0.0121(22) | +0.0733(10) | 23.952(49) | 23.817(34) |
| 2009/12/12  7.3791 | Keck/NIRC2 | 42.853 | 41.872 | 0.11 | −0.0244(30) | −0.0672(30) | - | - |
| **(148780) Altjira** | | | | | | | | |
| 2006/08/06  6.3999 | ACS | 45.353 | 45.623 | 1.23 | −0.1718(11) | +0.0630(41) | - | - |
| 2007/07/25  1.9308 | WFPC2 | 45.403 | 45.886 | 1.12 | −0.1633(16) | −0.0367(17) | 23.747(23) | 24.385(16) |
| 2007/08/07 22.3433 | WFPC2 | 45.405 | 45.672 | 1.23 | −0.2768(21) | −0.0074(10) | 23.658(27) | 23.737(32) |
| 2007/10/06 21.5919 | WFPC2 | 45.414 | 44.736 | 0.93 | −0.0528(10) | +0.0455(19) | 23.698(44) | 23.668(30) |
| 2007/11/12  9.2313 | WFPC2 | 45.419 | 44.449 | 0.25 | +0.1405(14) | −0.0632(12) | 23.52(10) | 23.960(85) |
| 2008/07/25  7.1686 | WFPC2 | 45.456 | 45.942 | 1.12 | +0.1572(13) | −0.0086(14) | 23.860(33) | 23.571(20) |
| 2009/12/11  9.4078 | Keck/NIRC2 | 45.528 | 44.590 | 0.38 | −0.3653(30) | +0.0379(30) | - | - |
| 2010/08/03 14.6229 | Keck/NIRC2 | 45.562 | 45.952 | 1.17 | −0.0359(60) | −0.0394(30) | - | - |
| **2001 QY$_{297}$** | | | | | | | | |
| 2006/04/18 17.2541 | ACS | 42.882 | 43.163 | 1.28 | +0.06581(50) | −0.05436(50) | - | - |
| 2007/08/17 13.8433 | WFPC2 | 42.988 | 41.987 | 0.20 | −0.4314(12) | −0.0082(10) | 23.388(86) | 23.048(14) |
| 2007/08/19  9.6392 | WFPC2 | 42.989 | 41.993 | 0.24 | −0.4407(18) | −0.0143(11) | 22.708(11) | 23.29(21) |
| 2008/04/11  1.8561 | WFPC2 | 43.040 | 43.476 | 1.20 | −0.0551(13) | +0.0786(11) | 23.447(52) | 23.150(30) |
| 2008/04/30 21.2170 | WFPC2 | 43.045 | 43.160 | 1.33 | −0.3022(10) | +0.0398(10) | 23.23(15) | 23.43(21) |
| 2008/08/01  9.4436 | WFPC2 | 43.065 | 42.059 | 0.18 | +0.1679(17) | −0.0081(12) | 23.038(98) | 23.210(95) |
| **2003 QW$_{111}$** | | | | | | | | |
| 2006/07/25  9.1349 | ACS | 44.743 | 43.949 | 0.82 | +0.2948(10) | −0.1348(14) | - | - |
| 2007/07/25  3.6392 | WFPC2 | 44.643 | 43.868 | 0.85 | +0.1554(22) | −0.0975(10) | 24.43(10) | 24.980(40) |
| 2007/08/26 13.7017 | WFPC2 | 44.634 | 43.637 | 0.21 | +0.028(38) | −0.023(44) | 24.108(33) | 26.38(18) |
| 2008/08/04 19.2138 | WFPC2 | 44.539 | 43.666 | 0.67 | +0.1612(24) | −0.0504(27) | - | - |
| 2008/08/20 15.5804 | WFPC2 | 44.535 | 43.559 | 0.35 | +0.2622(16) | −0.1096(13) | 23.89(51) | 25.31(13) |
| 2008/09/07 14.0936 | WFPC2 | 44.530 | 43.524 | 0.08 | +0.3004(40) | −0.1347(40) | - | - |
| 2008/10/26 19.5369 | WFPC2 | 44.516 | 43.893 | 1.00 | −0.0759(64) | +0.0208(38) | - | - |
| **2003 QY$_{90}$** | | | | | | | | |
| 2003/10/23  1.2142 | Lit:KE'06 | 44.973 | 44.557 | 1.16 | +0.336(26) | −0.235(32) | - | - |
| 2004/09/13  2.8685 | Lit:KE'06 | 44.973 | 44.065 | 0.56 | +0.329(13) | −0.190(18) | - | - |
| 2005/06/08  9.4375 | Lit:KE'06 | 44.973 | 44.615 | 1.22 | +0.2321(99) | −0.1566(98) | - | - |
| 2005/06/09  8.1182 | Lit:KE'06 | 44.973 | 44.600 | 1.21 | +0.238(20) | −0.166(20) | - | - |
| 2005/07/09  7.9363 | Lit:KE'06 | 44.973 | 44.192 | 0.84 | +0.318(13) | −0.169(17) | - | - |
| 2005/07/25  2.2740 | ACS | 44.973 | 44.050 | 0.55 | +0.32118(58) | −0.1700(25) | 24.124(36) | 24.14(12) |
| 2005/08/09  4.6944 | Lit:K'05 | 44.973 | 43.976 | 0.23 | +0.3188(44) | −0.1420(58) | - | - |
| 2005/08/14 23.2042 | ACS | 44.973 | 43.964 | 0.12 | +0.3217(11) | −0.15645(64) | 24.134(53) | 24.02(32) |
| 2005/09/04 11.7314 | ACS | 44.973 | 44.002 | 0.35 | +0.3076(15) | −0.1380(11) | 24.122(19) | 24.53(11) |
| 2006/09/17 21.0077 | ACS | 44.972 | 44.079 | 0.59 | +0.1905(11) | −0.0402(18) | 24.106(19) | 24.28(12) |
| 2007/10/11  9.4519 | WFPC2 | 44.971 | 44.317 | 0.97 | −0.0644(15) | +0.0755(18) | 24.395(36) | 24.305(25) |
| 2009/05/10  2.1464 | WFPC2 | 44.966 | 45.177 | 1.26 | +0.1044(10) | +0.0312(39) | - | - |

Table notes:

[a.] Unless otherwise indicated, the camera used with ACS was the HRC, the camera used with WFPC2 was the PC, and the camera used with NICMOS was NIC2. Astrometry from the literature is indicated by "Lit:O'03" for Osip et al. (2003), "Lit:K'05" for Kern (2005), and "Lit:KE'06" for Kern and Elliot (2006).

[b.] The distance from the Sun to the target is $r$ and from the observer to the target is $\Delta$. The phase angle is the angular separation between the observer and Sun as seen from the target.

[c.] Relative right ascension $\Delta x$ and relative declination $\Delta y$ are computed as $\Delta x = (\alpha_2 - \alpha_1)\cos(\delta_1)$ and $\Delta y = \delta_2 - \delta_1$, where $\alpha$ is right ascension, $\delta$ is declination, and subscripts 1 and 2 refer to primary and secondary, respectively. Estimated 1-$\sigma$ uncertainties in the final digits are indicated in parentheses. For HST observations, uncertainties are estimated from the scatter between fits to individual frames. For Keck, the uncertainty estimate is inflated slightly to account for potential systematic error sources we are not aware of.

[d.] Visual photometry for primary and secondary bodies is reported, where available. For HST observations, this meant visits with sufficient spatial separation to extract separate photometry for primary and secondary in observations that used $V$-like filters (specifically *F475W*, *F555W*, or *F606W*, converted to $V$ using synphot as described in detail by Benecchi et al. 2009). The Keck observations were all done in $H$ band, without photometric standard

stars, so no photometry is reported. Estimated 1-σ uncertainties in the final digits are indicated in parentheses. Although many of the published astrometric observations do include photometry, they are not in *V*-like filters, so those results are not duplicated here.

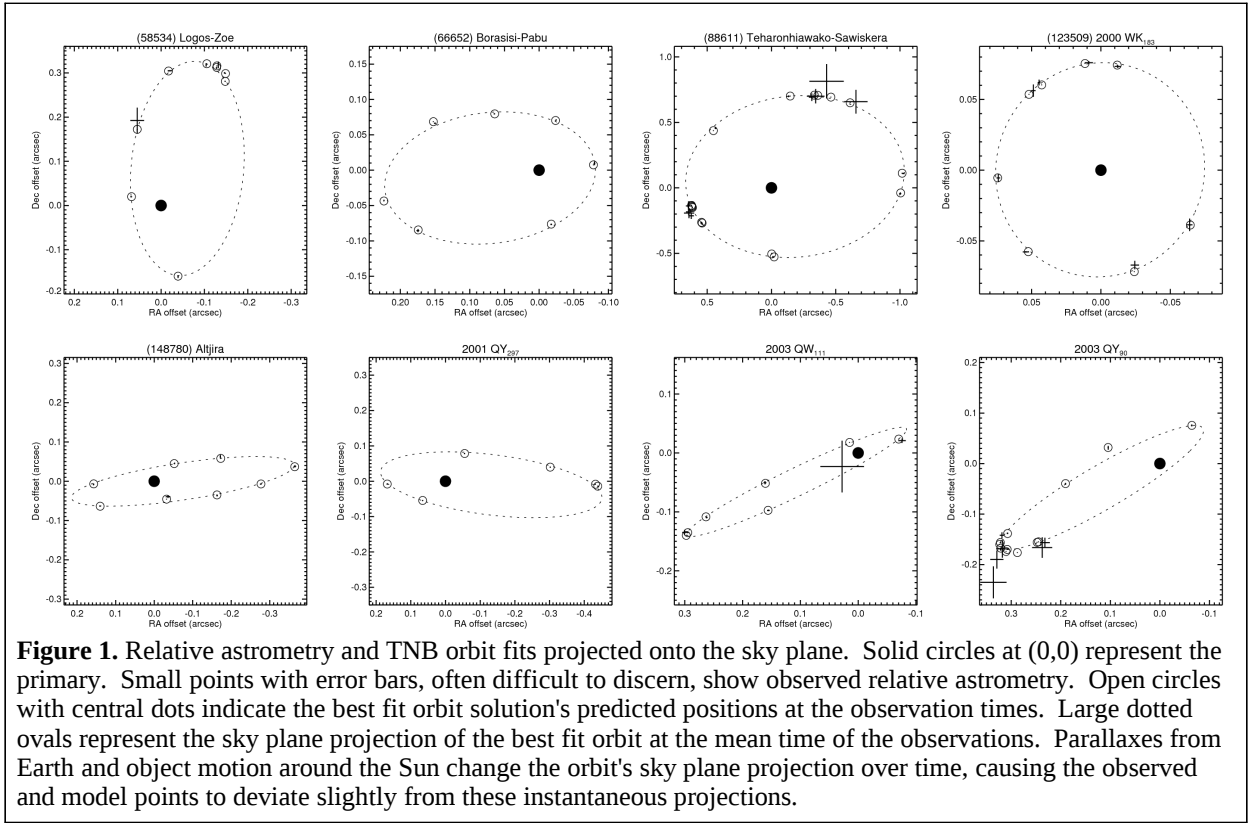

**Figure 1.** Relative astrometry and TNB orbit fits projected onto the sky plane. Solid circles at (0,0) represent the primary. Small points with error bars, often difficult to discern, show observed relative astrometry. Open circles with central dots indicate the best fit orbit solution's predicted positions at the observation times. Large dotted ovals represent the sky plane projection of the best fit orbit at the mean time of the observations. Parallaxes from Earth and object motion around the Sun change the orbit's sky plane projection over time, causing the observed and model points to deviate slightly from these instantaneous projections.

For each system, Keplerian orbits were fitted to the astrometric data and uncertainties using nonlinear least squares minimization procedures described by Grundy et al. (2009). Astrometric errors were assumed to obey Gaussian distributions. Where possible, optimal scheduling techniques (Grundy et al. 2008) were used to time subsequent observations so as to minimize the number required to obtain a definitive orbit solution. It is worth describing here what we mean when we mean by a definitive TNB orbit solution. Our arbitrary criteria for orbit knowledge are that fractional uncertainties in the period and semimajor axis must be less than 5%, eccentricity uncertainty must be less than 0.05, and the cumulative period uncertainty after two years must be less than 10% of one orbital period. In some systems, the similar brightnesses of primary and secondary make their identities in subsequent visits uncertain until enough astrometric information is available to simultaneously fit the orbit and exclude all other identity permutations. The new orbit solutions resulting from these fits are listed in Tables 3-10, along with $\chi^2$. Sky-plane astrometry and instantaneous projections of the new orbits are illustrated in Fig. 1. Sky-plane residuals for relative astrometry from ACS/HRC and WFPC2/PC images are typically in the milliarcsec range, as expected from the estimated astrometric uncertainties.

**Table 3**
Improved orbital parameters and 1-σ uncertainties for (58534) Logos-Zoe.

| Parameter | | Value |
|---|---|---|
| Fitted elements:[a] | | |
| Period (days) | $P$ | $309.87 \pm 0.22$ |
| Semimajor axis (km) | $a$ | $8217 \pm 42$ |
| Eccentricity | $e$ | $0.5463 \pm 0.0079$ |
| Inclination[b] (deg) | $i$ | $95.43 \pm 0.47$ |
| Mean longitude[b] at epoch[c] (deg) | $\epsilon$ | $259.47 \pm 0.66$ |
| Longitude of asc. node[b] (deg) | $\Omega$ | $16.07 \pm 0.34$ |
| Longitude of periapsis[b] (deg) | $\varpi$ | $223.80 \pm 0.76$ |
| Derived parameters: | | |
| Standard gravitational parameter $GM_{sys}$ (km$^3$ day$^{-2}$) | $\mu$ | $0.03056 \pm 0.00047$ |
| System mass ($10^{18}$ kg) | $M_{sys}$ | $0.4579 \pm 0.0069$ |
| Orbit pole right ascension[b] (deg) | $\alpha_{pole}$ | $286.07 \pm 0.35$ |
| Orbit pole declination[b] (deg) | $\delta_{pole}$ | $-5.43 \pm 0.47$ |
| Orbit pole ecliptic longitude[d] (deg) | $\lambda_{pole}$ | $286.75 \pm 0.35$ |
| Orbit pole ecliptic latitude[d] (deg) | $\beta_{pole}$ | $+17.08 \pm 0.48$ |
| Next mutual events season | | 2028 |

Table notes:

[a] Elements are for secondary relative to primary. Excluding the initial observation made with the lower resolution WFPC2/WF3 camera in 2001, the average sky plane residual for Orbit 1 is 2.2 mas and the maximum is 3.9 mas; $\chi^2$ is 20, based on observations at 9 epochs. The mirror orbit solution has an average residual of 13 mas and $\chi^2$ of 478 and is formally excluded if we assume Gaussian errors and independent observations.

[b] Referenced to J2000 equatorial frame.

[c] The epoch is Julian date 2452600.0 (2002 November 21 12:00 UT).

[d] Referenced to J2000 ecliptic frame.

**Table 4**
Improved orbital parameters and 1-σ uncertainties for (66652) Borasisi-Pabu.

| Parameter | | Value |
|---|---|---|
| Fitted elements:[a] | | |
|    Period (days) | $P$ | $46.2888 \pm 0.0018$ |
|    Semimajor axis (km) | $a$ | $4528 \pm 12$ |
|    Eccentricity | $e$ | $0.4700 \pm 0.0018$ |
|    Inclination[b] (deg) | $i$ | $54.31 \pm 0.30$ |
|    Mean longitude[b] at epoch[c] (deg) | $\epsilon$ | $214.40 \pm 0.50$ |
|    Longitude of asc. node[b] (deg) | $\Omega$ | $70.44 \pm 0.47$ |
|    Longitude of periapsis[b] (deg) | $\varpi$ | $238.85 \pm 0.66$ |
| Derived parameters: | | |
|    Standard gravitational parameter $GM_{sys}$ (km$^3$ day$^{-2}$) | $\mu$ | $0.2291 \pm 0.0018$ |
|    System mass ($10^{18}$ kg) | $M_{sys}$ | $3.433 \pm 0.027$ |
|    Orbit pole right ascension[b] (deg) | $\alpha_{pole}$ | $340.44 \pm 0.47$ |
|    Orbit pole declination[b] (deg) | $\delta_{pole}$ | $+35.69 \pm 0.30$ |
|    Orbit pole ecliptic longitude[d] (deg) | $\lambda_{pole}$ | $358.70 \pm 0.47$ |
|    Orbit pole ecliptic latitude[d] (deg) | $\beta_{pole}$ | $+40.05 \pm 0.33$ |
|    Next mutual events season | | 2078 |

Table notes:

[a] Elements are for secondary relative to primary. The average sky plane residual for Orbit 1 is 1.4 mas and the maximum is 3.3 mas; $\chi^2$ is 17, based on observations at 7 epochs. The mirror orbit solution has an average residual of 4.4 mas and $\chi^2$ of 148 and is formally excluded if we assume Gaussian errors and independent observations.

[b] Referenced to J2000 equatorial frame.

[c] The epoch is Julian date 2451900.0 (2000 December 21 12:00 UT).

[d] Referenced to J2000 ecliptic frame.

**Table 5**

Improved orbital parameters and 1-σ uncertainties for (88611) Teharonhiawako-Sawiskera.

| Parameter | | Value |
|---|---|---|
| Fitted elements:[a] | | |
| Period (days) | $P$ | $828.76 \pm 0.22$ |
| Semimajor axis (km) | $a$ | $27670 \pm 120$ |
| Eccentricity | $e$ | $0.2494 \pm 0.0021$ |
| Inclination[b] (deg) | $i$ | $144.42 \pm 0.35$ |
| Mean longitude[b] at epoch[c] (deg) | $\epsilon$ | $296.2 \pm 1.1$ |
| Longitude of asc. node[b] (deg) | $\Omega$ | $54.22 \pm 0.69$ |
| Longitude of periapsis[b] (deg) | $\varpi$ | $20.1 \pm 1.0$ |
| Derived parameters: | | |
| Standard gravitational parameter $GM_{sys}$ (km$^3$ day$^{-2}$) | $\mu$ | $0.1632 \pm 0.0021$ |
| System mass ($10^{18}$ kg) | $M_{sys}$ | $2.445 \pm 0.032$ |
| Orbit pole right ascension[b] (deg) | $\alpha_{pole}$ | $324.22 \pm 0.68$ |
| Orbit pole declination[b] (deg) | $\delta_{pole}$ | $-54.43 \pm 0.360$ |
| Orbit pole ecliptic longitude[d] (deg) | $\lambda_{pole}$ | $306.60 \pm 0.50$ |
| Orbit pole ecliptic latitude[d] (deg) | $\beta_{pole}$ | $-37.66 \pm 0.35$ |
| Next mutual events season | | 2061 |

Table notes:

[a.] Elements are for secondary relative to primary. Average sky plane residuals for our Keck observations are 6.4 mas; $\chi^2$ is 42, based on observations at 16 epochs. The mirror orbit solution has average Keck residuals of 20 mas and $\chi^2$ of 150 and is formally excluded if we assume Gaussian errors and independent observations.

[b.] Referenced to J2000 equatorial frame.

[c.] The epoch is Julian date 2452000.0 (2001 March 31 12:00 UT).

[d.] Referenced to J2000 ecliptic frame.

**Table 6**
Orbital parameters and 1-σ uncertainties for (123509) 2000 WK$_{183}$.

| Parameter | | Orbit 1 ($\chi^2 = 15$) | Orbit 2 ($\chi^2 = 15$) |
|---|---|---|---|
| Fitted elements:[a] | | | |
| Period (days) | $P$ | 30.9181 ± 0.0051 | 30.9159 ± 0.0047 |
| Semimajor axis (km) | $a$ | 2367 ± 27 | 2366 ± 28 |
| Eccentricity | $e$ | 0.0081 ± 0.0072 | 0.0086 ± 0.0078 |
| Inclination[b] (deg) | $i$ | 78.3 ± 6.0 | 59.8 ± 5.6 |
| Mean longitude[b] at epoch[c] (deg) | $\epsilon$ | 267.7 ± 3.5 | 281.5 ± 3.2 |
| Longitude of asc. node[b] (deg) | $\Omega$ | 152.4 ± 4.3 | 173.8 ± 5.8 |
| Longitude of periapsis[b] (deg) | $\varpi$ | 0 ± 75 | 29 ± 15 |
| Derived parameters: | | | |
| Standard gravitational parameter $GM_{sys}$ (km$^3$ day$^{-2}$) | $\mu$ | 0.0733 ± 0.0024 | 0.0733 ± 0.0026 |
| System mass (10$^{18}$ kg) | $M_{sys}$ | 1.099 ± 0.037 | 1.099 ± 0.039 |
| Orbit pole right ascension[b] (deg) | $\alpha_{pole}$ | 62.4 ± 3.8 | 83.8 ± 5.3 |
| Orbit pole declination[b] (deg) | $\delta_{pole}$ | +11.7 ± 4.9 | +30.2 ± 4.5 |
| Orbit pole ecliptic longitude[d] (deg) | $\lambda_{pole}$ | 62.7 ± 3.9 | 84.6 ± 4.6 |
| Orbit pole ecliptic latitude[d] (deg) | $\beta_{pole}$ | –9.2 ± 4.7 | +6.9 ± 4.6 |
| Next mutual events season | | 2067 | 2084 |

Table notes:

[a.] Elements are for secondary relative to primary. The average sky plane residual for Orbit 1 is 2.3 mas and the maximum is 4.6 mas; $\chi^2$ is 15, based on observations at 8 epochs. For Orbit 2 the average residual is also 2.3 mas.

[b.] Referenced to J2000 equatorial frame.

[c.] The epoch is Julian date 2454300.0 (2007 July 18 12:00 UT).

[d.] Referenced to J2000 ecliptic frame.

**Table 7**
Orbital parameters and 1-σ uncertainties for (148780) Altjira.

| Parameter | | Value |
|---|---|---|
| Fitted elements:[a] | | |
|    Period (days) | $P$ | 139.561 ± 0.047 |
|    Semimajor axis (km) | $a$ | 9904 ± 56 |
|    Eccentricity | $e$ | 0.3445 ± 0.0045 |
|    Inclination[b] (deg) | $i$ | 35.19 ± 0.19 |
|    Mean longitude[b] at epoch[c] (deg) | $\epsilon$ | 224.89 ± 0.16 |
|    Longitude of asc. node[b] (deg) | $\Omega$ | 312.73 ± 0.31 |
|    Longitude of periapsis[b] (deg) | $\varpi$ | 101.96 ± 0.91 |
| Derived parameters: | | |
|    Standard gravitational parameter $GM_{sys}$ (km³ day⁻²) | $\mu$ | 0.2638 ± 0.0045 |
|    System mass (10¹⁸ kg) | $M_{sys}$ | 3.952 ± 0.067 |
|    Orbit pole right ascension[b] (deg) | $\alpha_{pole}$ | 222.73 ± 0.30 |
|    Orbit pole declination[b] (deg) | $\delta_{pole}$ | +54.81 ± 0.19 |
|    Orbit pole ecliptic longitude[d] (deg) | $\lambda_{pole}$ | 184.56 ± 0.36 |
|    Orbit pole ecliptic latitude[d] (deg) | $\beta_{pole}$ | +64.87 ± 0.21 |
|    Next mutual events season | | 2028 |

Table notes:

[a.] Elements are for secondary relative to primary. The average sky plane residual for Orbit 1 is 2.4 mas and the maximum is 7.3 mas; $\chi^2$ is 11, based on observations at 8 epochs. The mirror orbit has an average residual of 6.5 mas and $\chi^2$ value of 65 and is formally excluded if we assume Gaussian errors and independent observations.

[b.] Referenced to J2000 equatorial frame.

[c.] The epoch is Julian date 2454300.0 (2007 July 18 12:00 UT).

[d.] Referenced to J2000 ecliptic frame.

**Table 8**
Orbital parameters and 1-σ uncertainties for 2001 QY$_{297}$.

| Parameter | | Value |
|---|---|---|
| Fitted elements:[a] | | |
| Period (days) | $P$ | 138.110 ± 0.023 |
| Semimajor axis (km) | $a$ | 9960 ± 31 |
| Eccentricity | $e$ | 0.4175 ± 0.0023 |
| Inclination[b] (deg) | $i$ | 172.86 ± 0.20 |
| Mean longitude[b] at epoch[c] (deg) | $\epsilon$ | 34.4 ± 2.6 |
| Longitude of asc. node[b] (deg) | $\Omega$ | 135.3 ± 1.3 |
| Longitude of periapsis[b] (deg) | $\varpi$ | 213.0 ± 2.8 |
| Derived parameters: | | |
| Standard gravitational parameter $GM_{sys}$ (km³ day⁻²) | $\mu$ | 0.2740 ± 0.0026 |
| System mass (10¹⁸ kg) | $M_{sys}$ | 4.105 ± 0.038 |
| Orbit pole right ascension[b] (deg) | $\alpha_{pole}$ | 45.3 ± 1.2 |
| Orbit pole declination[b] (deg) | $\delta_{pole}$ | −82.87 ± 0.20 |
| Orbit pole ecliptic longitude[d] (deg) | $\lambda_{pole}$ | 285.56 ± 0.69 |
| Orbit pole ecliptic latitude[d] (deg) | $\beta_{pole}$ | −71.00 ± 0.11 |
| Next mutual events season | | 2062 |

Table notes:

[a.] Elements are for secondary relative to primary. The average sky plane residual for this orbit solution is 0.6 mas and the maximum is 1.7 mas; $\chi^2$ is 1.9, based on observations at 6 epochs. The mirror orbit has an average residual of 2.0 mas and $\chi^2$ value of 28 and is formally excluded if we assume Gaussian errors and independent observations.

[b.] Referenced to J2000 equatorial frame.

[c.] The epoch is Julian date 2454200.0 (2007 April 9 12:00 UT).

[d.] Referenced to J2000 ecliptic frame.

**Table 9**
Orbital parameters and 1-σ uncertainties for 2003 QW$_{111}$.

| Parameter | | Orbit 1 ($\chi^2$ = 6.2) | Orbit 2 ($\chi^2$ = 16) |
|---|---|---|---|
| Fitted elements:[a] | | | |
| Period (days) | $P$ | 110.13 ± 0.12 | 110.11 ± 0.10 |
| Semimajor axis (km) | $a$ | 6681 ± 47 | 6760 ± 48 |
| Eccentricity | $e$ | 0.560 ± 0.012 | 0.547 ± 0.010 |
| Inclination[b] (deg) | $i$ | 25.46 ± 0.28 | 149.57 ± 0.29 |
| Mean longitude[b] at epoch[c] (deg) | $\epsilon$ | 126.46 ± 0.62 | 237.1 ± 1.1 |
| Longitude of asc. node[b] (deg) | $\Omega$ | 163.12 ± 0.87 | 312.84 ± 0.63 |
| Longitude of periapsis[b] (deg) | $\varpi$ | 251.7 ± 2.6 | 4.3 ± 2.4 |
| Derived parameters: | | | |
| Standard gravitational parameter $GM_{sys}$ (km$^3$ day$^{-2}$) | $\mu$ | 0.1300 ± 0.0028 | 0.1348 ± 0.0029 |
| System mass (10$^{18}$ kg) | $M_{sys}$ | 1.948 ± 0.041 | 2.019 ± 0.043 |
| Orbit pole right ascension[b] (deg) | $\alpha_{pole}$ | 73.11 ± 0.86 | 222.84 ± 0.63 |
| Orbit pole declination[b] (deg) | $\delta_{pole}$ | +64.54 ± 0.28 | –59.57 ± 0.30 |
| Orbit pole ecliptic longitude[d] (deg) | $\lambda_{pole}$ | 80.38 ± 0.48 | 240.59 ± 0.46 |
| Orbit pole ecliptic latitude[d] (deg) | $\beta_{pole}$ | +41.66 ± 0.29 | –40.85 ± 0.26 |
| Next mutual events season | | 2016 | 2125 |

Table notes:

[a.] Elements are for secondary relative to primary. The average sky plane residual for Orbit 1 is 2.7 mas and the maximum is 6.3 mas, excluding the 2007/08/26 observation when the separation was too close to resolve. $\chi^2$ is 6.2, based on observations at 7 epochs. For Orbit 2 the average residual is 3.9 mas.

[b.] Referenced to J2000 equatorial frame.

[c.] The epoch is Julian date 2454400.0 (2007 October 26 12:00 UT).

[d.] Referenced to J2000 ecliptic frame.

**Table 10**
Orbital parameters and 1-σ uncertainties for 2003 QY$_{90}$.

| Parameter | | Value |
|---|---|---|
| Fitted elements:[a] | | |
| Period (days) | $P$ | $309.68 \pm 0.18$ |
| Semimajor axis (km) | $a$ | $8549 \pm 95$ |
| Eccentricity | $e$ | $0.6625 \pm 0.0088$ |
| Inclination[b] (deg) | $i$ | $41.79 \pm 0.59$ |
| Mean longitude[b] at epoch[c] (deg) | $\epsilon$ | $313.2 \pm 1.4$ |
| Longitude of asc. node[b] (deg) | $\Omega$ | $113.57 \pm 0.39$ |
| Longitude of periapsis[b] (deg) | $\varpi$ | $286.7 \pm 1.3$ |
| Derived parameters: | | |
| Standard gravitational parameter $GM_{sys}$ (km$^3$ day$^{-2}$) | $\mu$ | $0.0345 \pm 0.0012$ |
| System mass ($10^{18}$ kg) | $M_{sys}$ | $0.516 \pm 0.017$ |
| Orbit pole right ascension[b] (deg) | $\alpha_{pole}$ | $23.57 \pm 0.39$ |
| Orbit pole declination[b] (deg) | $\delta_{pole}$ | $+48.21 \pm 0.60$ |
| Orbit pole ecliptic longitude[d] (deg) | $\lambda_{pole}$ | $41.54 \pm 0.55$ |
| Orbit pole ecliptic latitude[d] (deg) | $\beta_{pole}$ | $+35.32 \pm 0.49$ |
| Next mutual events season | | 2135 |

Table notes:

[a.] Elements are for secondary relative to primary. The average HST data sky plane residual for Orbit 1 is 0.8 mas and the maximum is 2.5 mas (residuals are much larger for some of the ground-based data from the literature included in the solution); $\chi^2$ is 25, based on observations at 12 epochs. The mirror orbit solution has an average residual of 7.9 mas and $\chi^2$ of 134 and is formally excluded if we assume Gaussian errors and independent observations.

[b.] Referenced to J2000 equatorial frame.

[c.] The epoch is Julian date 2453500.0 (2005 May 9 12:00 UT).

[d.] Referenced to J2000 ecliptic frame.

## 3. Properties Derived From Binary Mutual Orbits

With our new orbits we now have 22 TNB systems for which the period $P$, semimajor axis $a$, and eccentricity $e$ are known. For 13 of the 22, the existing observations permit two orbital solutions. These are mirror images of one another through the sky plane at the time of the observations, sharing the same or very similar $P$, $a$, and $e$. One is prograde and one is retrograde with respect to the TNB's heliocentric orbit (this is not absolutely required, but geometric circumstances for both solutions to be prograde or both be retrograde are unusual). We refer to these as "Orbit 1" and "Orbit 2" with Orbit 1 always taken as being the lower $\chi^2$ solution based on currently available data. Additional observations to formally exclude Orbit 2 at greater than 3-σ confidence are not yet available, pending accumulation of sufficient parallax due to Earth and TNB heliocentric motion. The $\chi^2$ value corresponding to 3-σ confidence is computed using the $\chi^2$ probability function as described by Press et al. (1992), assuming Gaussian errors and that each separate observation listed in Table 2 provides two independent constraints. For these ambiguous systems, we adopt $P$, $a$, and $e$ values intermediate between the two mirror solutions, with uncertainties inflated to encompass the full range of 1-σ uncertainties for the two solutions.

**Table 11**

Adopted orbital elements

| TNB system | Period $P$ (days)[a] | Semimajor axis $a$ (km)[a] | Eccentricity $e$[a] |
|---|---|---|---|
| This work | | | |
| (58534) Logos | 309.87 ± 0.22 | 8217 ± 42 | 0.5463 ± 0.0079 |
| (66652) Borasisi | 46.2888 ± 0.0018 | 4528 ± 12 | 0.4700 ± 0.0018 |
| (88611) Teharonhiawako | 828.76 ± 0.22 | 27670 ± 120 | 0.2494 ± 0.0021 |
| (123509) 2000 WK$_{183}$ | 30.9172 ± 0.0060 | 2366 ± 28 | 0.0086 ± 0.0078 |
| (148780) Altjira | 139.561 ± 0.047 | 9904 ± 56 | 0.3445 ± 0.0045 |
| 2001 QY$_{297}$ | 138.110 ± 0.023 | 9960 ± 31 | 0.4175 ± 0.0023 |
| 2003 QW$_{111}$ | 110.13 ± 0.12 | 6721 ± 88 | 0.554 ± 0.018 |
| 2003 QY$_{90}$ | 309.68 ± 0.18 | 8549 ± 95 | 0.6625 ± 0.0088 |
| Published orbits | | | |
| (26308) 1998 SM$_{165}$ | 130.1640 ± 0.0055 | 11366 ± 10 | 0.47337 ± 0.00077 |
| (42355) Typhon | 18.9709 ± 0.0063 | 1628 ± 30 | 0.526 ± 0.015 |
| (65489) Ceto | 9.554 ± 0.011 | 1840 ± 44 | 0.0043 ± 0.0087 |
| (90482) Orcus | 9.5388 ± 0.0011 | 9006 ± 16 | 0.0008 ± 0.0013 |
| (120347) Salacia | 5.49380 ± 0.00022 | 5619 ± 89 | 0.0084 ± 0.0076 |
| (134860) 2000 OJ$_{67}$ | 22.0412 ± 0.0040 | 2357 ± 41 | 0.088 ± 0.024 |
| (136199) Eris | 15.7872 ± 0.0014 | 37580 ± 260 | 0.0166 ± 0.0069 |
| 1998 WW$_{31}$ | 590 ± 38 | 22420 ± 900 | 0.823 ± 0.032 |
| 1999 OJ$_4$ | 84.114 ± 0.038 | 3267 ± 61 | 0.365 ± 0.013 |
| 2000 QL$_{251}$ | 56.450 ± 0.025 | 5002 ± 28 | 0.4870 ± 0.0064 |
| 2001 QC$_{298}$ | 19.2307 ± 0.0010 | 3813 ± 15 | 0.3351 ± 0.0024 |
| 2001 XR$_{254}$ | 125.579 ± 0.049 | 9310 ± 49 | 0.5561 ± 0.0046 |
| 2003 TJ$_{58}$ | 137.32 ± 0.18 | 3768 ± 85 | 0.528 ± 0.011 |
| 2004 PB$_{108}$ | 97.0203 ± 0.0075 | 10401 ± 84 | 0.4383 ± 0.0030 |

Table notes:

[a.] Values and uncertainties here are chosen such that symmetric error bars encompass the 1-σ uncertainties in cases where two mirror orbit solutions exist. In order to be as self-consistent as possible, elements for systems with orbits published or pending publication here result from re-analysis of the data, including re-reduction of image data in the HST archive, so the values and uncertainties here may differ somewhat from numbers in the original papers. Those are, for 26308 and 2001 QC$_{298}$, Margot et al. (2004); Typhon, Grundy et al. (2008); Ceto, Grundy et al. (2007); Orcus, Brown et al. (2010); Salacia, Stansberry et al. (2011); Eris, Brown and Schaller (2007); 1998 WW$_{31}$, Veillet et al. (2002); 134860, 1999 OJ$_4$, 2000 QL$_{251}$, 2001 XR$_{254}$, 2003 TJ$_{58}$, and 2004 PB$_{108}$, Grundy et al. (2009). For 1998 WW$_{31}$ in particular, our solution is based only on the publicly available HST data. Additional ground-based astrometric data were used by Veillet et al. (2002) but not tabulated in their paper. Partial information on the orbit of 2001 QW$_{322}$ was published by Petit et al. (2008), but that object is not included here because it does not meet our orbit knowledge criteria described in Section 2.

Adopted $P$, $a$, and $e$ values are listed in Table 11. Quantities derived from these values ap-pear in Table 12. Derived quantities include the system mass $M_{sys}$, computed using the CODATA 2006 value of the gravitational constant $G = 6.67428 \times 10^{-11}$ m$^3$ kg$^{-1}$ s$^{-2}$ (Mohr et al. 2008):

$$M_{sys} = \frac{4\pi^2 a^3}{P^2 G}. \qquad (1)$$

The Hill radius $r_H$ describes the approximate region in which the TNB system's gravity dominates over solar gravity over the entire heliocentric orbit:

$$r_H = a_\odot(1-e_\odot)\sqrt[3]{\frac{M_{sys}}{3 M_\odot}}, \qquad (2)$$

where $a_\odot$ and $e_\odot$ refer to the heliocentric orbit, rather than the binary mutual orbit, and $M_\odot$ is the mass of the Sun (Hamilton and Burns 1992). The ratio of the semimajor axis $a$ of the mutual orbit to the Hill radius $r_H$ gives an indication of the tightness of the binary, and its susceptibility to disruption by external perturbations.

More speculatively, we can use the magnitude difference between primary and secondary $\Delta_{mag}$ (computed from values in Table 2, or from the literature, as appropriate) in conjunction with a plausible bulk density $\rho$ (we assume the range from 0.5 to 2 g cm$^{-3}$) and the assumption of equal albedos for the primary and secondary (consistent with their highly correlated colors, e.g., Benecchi et al. 2009) to compute plausible radii $R_1$ and $R_2$, for primary and secondary, respectively:

$$R_1 = \sqrt[3]{\frac{3 M_{sys}}{4\pi\rho(1+10^{-\frac{3}{5}\Delta_{mag}})}}. \qquad (3)$$

$$R_2 = R_1 10^{-\frac{1}{5}\Delta_{mag}}$$

These sizes can, in turn, be combined with photometry to compute a range of plausible geometric albedos $A_P$. For Solar $V$ magnitude at 1 AU of $-26.741$ obtained by convolving the Rieke et al. (2008) Solar spectral energy distribution with the Johnson $V$ filter (Michael Mueller personal communication 2008; for the derivation, see Pravec and Harris 2007), the albedo of each component $j$ with radius $R_j$ in km and absolute $V$ magnitude $H_{Vj}$ is

$$A_{Pj} = \left(\frac{671}{R_j}\right)^2 10^{-\frac{2}{5}H_{Vj}}, \qquad (4)$$

with $H_{Vj}$ computable from the combined system $H_V$ and $\Delta_{mag}$ according to

$$H_{V1} = -2.5\log_{10}\left(\frac{10^{-\frac{2}{5}H_V}}{1+10^{-\frac{2}{5}\Delta_{mag}}}\right). \qquad (5)$$

$$H_{V2} = H_{V1} + \Delta_{mag}$$

The angular momentum vector of the TNB system $\vec{J}$ is composed of orbital and spin components: $\vec{J} = \vec{J}_{orb} + \vec{J}_{spin}$. The magnitude of this vector can be normalized to $J' = \sqrt{G M_{sys}^3 R_{eff}}$ (where $R_{eff}$ is the radius of a single, spherical body with the combined volume of the two components), to obtain a dimensionless specific angular momentum $J/J'$, potentially diagnostic of formation mechanism (e.g., Canup 2005). Not knowing anything about the spin states of either of the binary components, we report $J_{orb}/J'$. For radius ratio $\gamma = R_2/R_1$, assuming equal densities of the two component bodies, we estimate the primary and secondary masses and $J_{orb}$ as follows:

$$M_1 = \frac{M_{sys}}{1+\gamma^3},$$

$$M_2 = M_{sys} - M_1, \qquad (6)$$

$$J_{orb} = \frac{2\pi}{P} a^2 \sqrt{1-e^2} \frac{M_1 M_2}{M_{sys}}.$$

These derived quantities are listed in Table 12.

**Table 12**

Properties derived from adopted elements

| TNB system | System mass $M_{sys}$ ($10^{18}$ kg)[a] | Hill radius $r_H$ (km) | $a/r_H$ | Inclination[b] (°) | Adopted $\Delta_{mag}$ | Primary radius[c] $R_1$ (km) | V band albedo[c] $A_p$ | Specific angular momentum[c] $J_{orb}/J'$ |
|---|---|---|---|---|---|---|---|---|
| This work | | | | | | | | |
| (58534) Logos | 0.4579 ± 0.0069 | 250,000 | 0.033 | 74.2 | 0.45 | 32 - 50 | 0.11 - 0.28 | 2.6 - 3.3 |
| (66652) Borasisi | 3.433 ± 0.027 | 500,000 | 0.009 | 49.4 | 0.45 | 62 - 98 | 0.10 - 0.25 | 1.4 - 1.8 |
| (88611) Teharonhiawako | 2.445 ± 0.032 | 480,000 | 0.058 | 127.7 | 0.70 | 57 - 90 | 0.16 - 0.39 | 3.2 - 4.1 |
| (123509) 2000 WK$_{183}$ | 1.099 ± 0.039 | 360,000 | 0.007 | 97.5 or 81.2 | 0.11 | 41 - 65 | 0.10 - 0.26 | 1.3 - 1.7 |
| (148780) Altjira | 3.986 ± 0.067 | 540,000 | 0.018 | 25.4 | 0.23 | 64 - 100 | 0.06 - 0.14 | 2.2 - 2.8 |
| 2001 QY$_{297}$ | 4.105 ± 0.038 | 540,000 | 0.019 | 161.0 | 0.20 | 64 - 100 | 0.13 - 0.32 | 2.2 - 2.8 |
| 2003 QW$_{111}$ | 1.984 ± 0.078 | 400,000 | 0.017 | 48.9 or 131.2 | 1.14 | 55 - 87 | 0.08 - 0.19 | 1.1 - 1.5 |
| 2003 QY$_{90}$ | 0.516 ± 0.017 | 270,000 | 0.032 | 51.4 | 0.03 | 31 - 50 | 0.18 - 0.44 | 2.9 - 3.8 |
| Orbits previously published | | | | | | | | |
| (26308) 1998 SM$_{165}$ | 6.867 ± 0.018 | 470,000 | 0.024 | 75.4 | 2.69 | 89 - 140 | 0.06 - 0.15 | 0.19 - 0.24 |
| (42355) Typhon | 0.949 ± 0.052 | 140,000 | 0.011 | 50.5 | 1.30 | 43 - 69 | 0.05 - 0.13 | 0.46 - 0.62 |
| (65489) Ceto | 5.41 ± 0.39 | 260,000 | 0.007 | 66.3 or 115.4 | 0.58 | 73 - 116 | 0.05 - 0.13 | 0.42 - 0.59 |
| (90482) Orcus | 636.1 ± 3.3 | 2,100,000 | 0.004 | 73.7 or 106.6 | 2.61 | 400 - 640 | 0.10 - 0.26 | 0.09 - 0.12 |
| (120347) Salacia | 466 ± 22 | 2,400,000 | 0.002 | 137.2 or 42.4 | 2.32 | 360 - 570 | 0.01 - 0.03 | 0.11 - 0.14 |
| (134860) 2000 OJ$_{67}$ | 2.14 ± 0.11 | 450,000 | 0.005 | 85.1 or 94.2 | 0.56 | 53 - 85 | 0.09 - 0.23 | 1.0 - 1.4 |
| (136199) Eris | 16880 ± 350 | 8,000,000 | 0.005 | 101.2 or 78.6 | 6.70 | 1260 - 2000 | 0.31 - 0.79 | 0.0004 - 0.0005 |
| 1998 WW$_{31}$ | 2.57 ± 0.31 | 460,000 | 0.049 | 51.9 or 128.0 | 0.40 | 56 - 89 | 0.09 - 0.23 | 3.1 - 4.6 |
| 1999 OJ$_4$ | 0.390 ± 0.022 | 230,000 | 0.014 | 57.6 or 119.4 | 0.09 | 29 - 46 | 0.13 - 0.32 | 1.8 - 2.5 |
| 2000 QL$_{251}$ | 3.112 ± 0.052 | 460,000 | 0.011 | 134.2 or 45.3 | 0.12 | 58 - 92 | 0.04 - 0.10 | 1.6 - 2.0 |
| 2001 QC$_{298}$ | 11.88 ± 0.14 | 760,000 | 0.005 | 73.5 | 0.44 | 93 - 150 | 0.03 - 0.07 | 1.0 - 1.3 |
| 2001 XR$_{254}$ | 4.055 ± 0.065 | 550,000 | 0.017 | 21.1 or 158.3 | 0.43 | 65 - 104 | 0.09 - 0.23 | 2.0 - 2.5 |
| 2003 TJ$_{58}$ | 0.225 ± 0.015 | 200,000 | 0.019 | 61.9 or 116.6 | 0.51 | 25 - 40 | 0.17 - 0.42 | 1.9 - 2.6 |
| 2004 PB$_{108}$ | 9.47 ± 0.23 | 700,000 | 0.015 | 83.2 or 95.5 | 1.32 | 93 - 150 | 0.02 - 0.05 | 0.9 - 1.2 |

Table notes:

a. System masses $M_{sys}$ are based on the CODATA 2006 value of the gravitational constant $G = 6.6742 \times 10^{-11}$ m$^3$ s$^{-2}$ kg$^{-1}$ (Mohr et al. 2008). As noted before, our re-analysis of existing data has led to slightly different values for some of the binaries compared with previously published orbits (see notes to Table 11 for specific references).

b. The inclination between the binary's mutual orbit plane and its heliocentric orbit plane. For systems where the mirror ambiguity remains unresolved, two angles are given. Inclinations above 90° indicate mutual orbits that are retrograde relative to the heliocentric orbit.

c. These ranges of values are based on the assumption of spherical shapes, densities from 0.5 to 2.0 g cm$^{-3}$, and equal albedos of primary and secondary, so that their relative sizes can be computed from their relative magnitudes, as described in the text. Better estimates of these parameters based on other sorts of data may exist for a few of these objects. The specific angular momentum $J_{orb}/J'$ includes only the orbital component, since the spin states of the bodies are unknown.

As mentioned above, crude albedo estimates can be made from system masses in conjunction with photometry and assumed densities. The five new orbits provide five new albedo estimates, as shown in Table 12. The lowest of these albedos is for 2003 QW$_{111}$, the only one of these five systems in a mean motion resonance with Neptune. The other four are Classical objects. The two highest albedos among them are for 2001 QY$_{297}$ and 2003 QY$_{90}$, which are also the two with the lowest mean inclinations relative to the plane of the Solar System. This albedo pattern is consistent with an earlier finding that the lowest inclination Classical objects have higher albedos than other, more excited classes of small TNOs (Grundy et al. 2005; Brucker et al. 2009).

It is worth noting that a few of the systems could undergo eclipse and/or occultation type mutual events observable from Earth during the next couple of decades. Such events offer opportunities to accurately determine the sizes of the bodies in the system, and thus their densities, as well as providing extremely precise orbital timing information. The Logos-Zoe system has mutual events coming up in the 2027-2028 time frame, but given the small sizes of Logos and Zoe (Noll et al. 2004a estimated $R_1$ and $R_2$ to be 40 and 33 km, respectively) and their long mutual orbital period of 309 days, few events are likely to be observable from Earth. More promising are our solutions for two other TNBs with shorter periods and somewhat larger components. The Altjira system has at least a dozen observable events during the 2025-2031 time frame and the Orbit 1 solution for 2003 QW$_{111}$ has a comparable number of events even earlier, during 2015-2017. Of course, that system still needs additional observations to determine whether or not its Orbit 1 solution is indeed the correct solution.

## 4. Ensemble Results

This sample of 22 TNB mutual orbits offers an opportunity to investigate the statistical characteristics of a fascinating population of planetesimals remaining from the outer parts of the protoplanetary nebula. However, it is important to recognize potential biases in the sample. Two main classes of biases exist: discovery biases and orbit determination biases. It is harder to recognize and discover closer binaries than well-separated ones, so the inventory of known close binaries is truncated by the spatial resolution limits of available telescopes (e.g., Noll et al. 2008a). Binary searches that visit each target only a single time may also miss systems which just happen to have small separations at the time of the observation because they are near periapsis or viewed edge-on with the secondary near conjunction. From Earth, TNB systems with low inclinations between the satellite orbit and the heliocentric orbit are always viewed edge-on. Large brightness differences between primaries and secondaries are also a factor in discovery efficiency, with faint secondaries potentially being lost in the glare of the primary.

Once a TNB system has been discovered to be a binary, additional biases affect the ability to determine its mutual orbit. When primary and secondary have near equal brightness, as many do, it can be difficult to distinguish which is which from one epoch to the next, requiring additional observations to sort out this ambiguity. Systems with very short periods (a few days) or very long periods (a few years) can be more difficult to handle, owing to mismatches with observing cadences available from the few telescopes capable of resolving them (e.g., Petit et al. 2008). For periods near one year, it is challenging to sample all orbital longitudes because successive apparitions coincide with the same orbital longitudes. Systems with high eccentricity spend most of their time near apoapsis, making it challenging to observe other orbital longitudes, and edge-on systems introduce ambiguities as to which side of the orbit the secondary is on in any given observation. These kinds of difficulties can all be overcome with sufficient observations, but they mean that the mutual orbits of these systems tend to be determined later than those of other, more conveniently configured systems. At present, we are aware of about two dozen TNB systems for which additional, resolved observations have already been made beyond the discovery epoch, but the orbit still remains to be determined. Securing their orbits will more than double the sample of known TNB orbits, and is the aim of a new 3-year NOAO (National Optical Astronomy Observatory) Survey program of Gemini laser guide star adaptive optics observations beginning in 2011.

**4.1 Orientations**

Schlichting and Sari (2008) discuss TNB orbit orientations in light of two possible formation scenarios described by Goldreich et al. (2002). These are the "$L^2s$" mechanism, in which dynamical friction from a sea of small bodies gradually damps the orbit of a transiently captured satellite, and the "$L^3$" mechanism, in which one additional comparably-sized body carries off the excess angular momentum. Schlichting and Sari's simulations show that the $L^2s$ mechanism should produce mostly retrograde binary systems, in contrast with the $L^3$ mechanism, which produces roughly equal numbers of prograde and retrograde systems. Our sample has nine TNB systems for which all orbital elements ($P$, $a$, $e$, $i$, $\epsilon$, $\Omega$, and $\varpi$) are known, of which seven are prograde and two are retrograde with respect to their heliocentric orbits (see Table 12). To these we could add the wide binary 2001 QW$_{322}$, which is known to orbit in the retrograde sense, although $P$, $a$, and $e$ remain uncertain (Petit et al. 2008). The resulting tally of seven prograde and three retrograde orbits is clearly inconsistent with the preponderance of retrograde orbits predicted if $L^2s$ is the dominant formation mechanism. It is more consistent with a random distribution of orbit orientations. For randomly oriented orbits, one would expect to draw three or fewer retrograde orbits in a sample of ten about 17% of the time. The statistics of orbit orientations is discussed in greater detail by Naoz et al. (2010). While the current data may favor exclusion of $L^2s$ as the sole binary formation mechanism, numerous other mechanisms have been proposed (e.g., Weidenschilling 2002; Funato et al. 2004; Astakhov et al. 2005; Farrelly et al. 2006; Lee et al. 2007; Noll et al. 2008a; Nesvorný et al. 2010) and several more binary orbits have recently been announced (e.g., Parker and Kavelaars 2010) so the prospects are good for improvement in the statistics of TNB orbit orientations, as will be needed to distinguish among multiple possible formation mechanisms.

**4.2 Eccentricities, Tidal Dissipation, and *Q***

The TNBs in our sample have a wide range of eccentricities, from the nearly-circular orbit of the Orcus-Vanth system to $e = 0.82 \pm 0.03$ for 1998 WW$_{31}$. Six systems have eccentricities below 0.1. A larger group of fifteen have intermediate eccentricities clustered between 0.25 and

0.66.

Given this sample of orbital eccentricities, an obvious question to ask is whether or not the systems with circular orbits today had initially more eccentric orbits that were subsequently tidally circularized. If so, their characteristics could perhaps be used to constrain internal dissipation in the component bodies. A crude estimate of the timescale for tidally damping a binary's eccentricity can be made by considering dissipation in a synchronously rotating satellite. The assumption of synchronous rotation is justified by the fact that despinning is always much faster than damping of eccentricity (however, finding a satellite in non-synchronous rotation could potentially provide a useful upper bound on how dissipative it might be). For a low orbital eccentricity, the damping timescale $\tau$ is given by

$$\tau = \frac{4}{63}\frac{M_2}{M_1}\frac{a^5}{R_2^5}\frac{Q_2\mu'}{n} \tag{7}$$

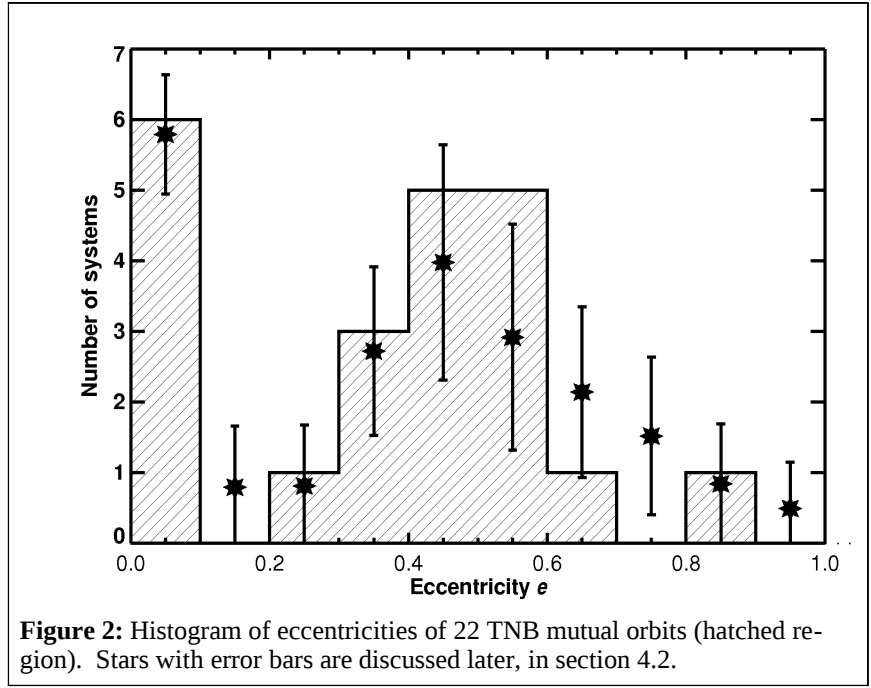

**Figure 2:** Histogram of eccentricities of 22 TNB mutual orbits (hatched region). Stars with error bars are discussed later, in section 4.2.

(Murray and Dermott 1999), where $M_1$ and $M_2$ are the masses of primary and secondary, $a$ is the mutual orbit's semimajor axis, $n$ is its mean motion, $R_2$ is the secondary's radius, $Q_2$ is its dissipation factor (e.g., Goldreich and Soter 1966), and $\mu'$ is a dimensionless measure of its rigidity. For a uniform satellite, $\mu'$ is given by $19\mu/2\rho_2 g_2 R_2$, where $\mu$ is the rigidity, $\rho_2$ is the density of the secondary, and $g_2$ is the gravitational acceleration at its surface due to $M_2$. We may rewrite equation (7) as follows:

$$\tau = \left[\frac{19}{42}\left(\frac{4}{3}\right)^{7/3}\frac{Q_2\mu\rho_2^{1/3}\pi^{4/3}}{G^{3/2}}\right]\frac{a^{13/2}(1+\gamma^3)^{7/3}}{M_{sys}^{17/6}\gamma^4} = \frac{1}{C}\frac{1}{X}. \tag{8}$$

Here $M_{sys} = M_1 + M_2$, the term inside the square brackets is denoted by $1/C$, the part outside is $1/X$, and we assume the densities of the primary and secondary are equal. For an initial eccentricity $e_0$, the eccentricity at time $t$ is given by

$$e = e_0 e^{-t/\tau} = e_0 e^{-CtX}. \tag{9}$$

Figure 3 shows a log-log plot of the observed eccentricity $e$ against the quantity $X$ for each TNB system (with $X$ in MKS units). The curves are from equation (9) with two different values of $Ct$, as indicated. Although there are few points and considerable scatter, values of $Ct$ in the range of $10^{-8}$ to $10^{-11}$ are reasonably consistent with the observations. From equation (8), if we set $t = 4.5$ Gyr, $\rho_2 = 1$ g cm$^{-3}$, and $\mu = 10$ Gpa, we obtain $Q_2$ ranging from 0.02 to 20, for $Ct$ from $10^{-8}$ to $10^{-11}$, respectively. A value of $Q_2$ of 0.02 is absurdly small. This order of magnitude range would imply significant dissipation for initially large eccentricities to have become damped in these systems, at least if this simple analysis were complete. But we have neglected effects of tides raised on the primary by the secondary as well as effects of Kozai oscillations (described in section 4.3) enhancing dissipation during episodes of maximum eccentricity, as proposed by Perets and Naoz (2009). A more complete analysis is the subject of a future paper. We stress that if some TNB systems formed with initially circular orbits, the key premise of this section collapses. It may also be unreasonable to expect both large and small TNOs to share similar interior structures and dissipation characteristics. With more data it may be possible to detect such differences.

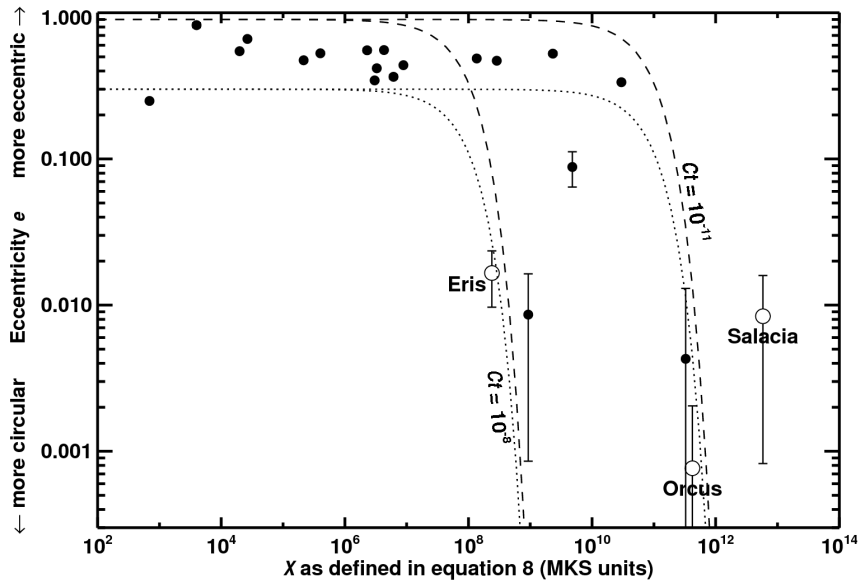

**Figure 3.** Log-log plot of eccentricity $e$ versus the parameter $X$, which depends on a number of assumptions and uncertain parameters as discussed in the text. For constant dissipation factor and rigidity, systems with larger $X$ should circularize more quickly. All low eccentricity systems in our sample have large $X$. Systems with masses larger than $10^{20}$ kg and primary radii larger than 300 km are depicted with open, labeled symbols. These systems are likely to have interior structures distinct from the more typical, smaller systems indicated by filled symbols. The curves are from Eq. 9, for initial eccentricities of 0.3 (dotted) and 0.9 (dashed).

### 4.3 Kozai Oscillations

Solar tides cause coupled oscillations of $e$ and $i$ in binary mutual orbits that are inclined with respect to their heliocentric orbits (often referred to as the Kozai mechanism, e.g., Kozai 1962; Nesvorný et al. 2003; Perets and Naoz 2009). Although orientations of many TNB orbits remain ambiguous as to which of the two mirror solutions is correct, we can already say that most of the 22 systems have inclinations relative to their heliocentric orbits in the range ($40° < i < 140°$) where such Kozai oscillations could be strong (see Table 12). Only 3 have inclinations outside this range (Altjira, 2001 QY$_{297}$, and 2001 XR$_{254}$). For 22 randomly oriented orbits, one would expect to count 5.1 ± 2.0 outside that range, consistent with the low observed number. Many of the TNB systems are evidently quite highly inclined with respect to their heliocentric orbital planes. Eight of the 22 (Eris, Logos, Orcus, 26308, 123509, 134860, 2001 QC$_{298}$, and 2004 PB$_{108}$) are within 20° of perpendicular to their heliocentric orbit planes. For random orientations, we would expect to observe 7.5 ± 2.2 out of 22 within 20° of perpendicular. Our data thus show no evidence for high inclination systems being scarcer than would be expected of a random distribution, as would occur if the Kozai mechanism coupled with tidal evolution caused the destruction of high-inclination systems (e.g., Perets and Naoz 2009).

One possible reason for the Kozai mechanism not to have destroyed high inclination TNBs is that bodies with irregular shapes have quadrupole and higher order terms in their gravitational fields. For tight binaries, effects of these terms can dominate over effects of solar tides (e.g.,

Fabrycky and Tremaine 2007). Nicholson et al. (2008) give the critical transition semimajor axis $a_c$ between oblateness-dominated and solar-tide-dominated dynamics as

$$a_c = \sqrt[5]{2 J_2 \frac{M_{sys}}{M_\odot} R_1^2 a_\odot^3}, \qquad (10)$$

where $J_2$ is the primary's second gravitational moment. For plausible values of $J_2$ between 0.001 and 0.2, we compared $a$ with $a_c$, finding that many of the TNBs could be in the oblateness-dominated regime.

For the systems where there was at least a chance of $a$ being larger than $a_c$, we ran n-body orbital integrations (including the Sun plus the TNB pair) to see how the Kozai oscillations might affect the binary orbits. These simulations used a conservative-force Bulirsch-Stoer integrator adapted from Mercury6 (MDT_BS2.FOR; Chambers 1999) with normalized spatial and velocity accuracies per time step of $10^{-12}$. The binary objects were initialized to state vectors based on the mutual orbits reported here, along with JPL Horizons ephemerides for their heliocentric orbits. After initialization, all objects were allowed to mutually gravitate for several Kozai oscillations, with instantaneous orbital elements recovered and output periodically. The observed cycles emerge from the tidal gravitation of the Sun, and are not explicit in the program itself. All systems (along with mirror orbits, where appropriate) show oscillations in $e$ and $i$, with minimum and maximum eccentricities and inclinations listed in Table 13, along with the plausible range for $a/a_c$. As expected, the oscillations have shorter periods for systems with longer period mutual orbits and/or more eccentric heliocentric orbits, and smaller amplitudes for systems with lower inclinations between heliocentric and mutual orbits.

**Table 13**
Kozai Oscillations from n-body Integrations

| TNB system (orbit) | Eccentricity excursion | Inclination excursion (°) | Kozai period[a] (years) | Precession period[a] (years) | $a/a_c$ ratio[b] |
|---|---|---|---|---|---|
| This work | | | | | |
| (58534) Logos | 0.43 – 0.96 | 34 – 75 | 37,000 | 81,000 | 0.9 – 3.0 |
| (66652) Borasisi | 0.29 – 0.71 | 35 – 53 | 300,000 | 670,000 | 0.3 – 0.9 |
| (88611) Teharonhiawako | 0.16 – 0.64 | 140 – 127 | 27,000 | 49,000 | 1.7 – 5.9 |
| (123509) 2000 WK$_{183}$ (1) | 0.00 – 0.98 | 141 – 99 | 1,100,000 | 1,300,000 | 0.2 – 0.7 |
| (123509) 2000 WK$_{183}$ (2) | 0.01 – 0.98 | 39 – 80 | 1,000,000 | 1,300,000 | 0.2 – 0.7 |
| (148780) Altjira (1) | 0.33 – 0.41 | 22 – 26 | 40,000 | 220,000 | 0.5 – 1.8 |
| (148780) Altjira (2) | 0.34 – 0.42 | 158 – 154 | 40,000 | 230,000 | 0.5 – 1.8 |
| 2001 QY$_{297}$ | 0.39 – 0.44 | 163 – 159 | 35,000 | 190,000 | 0.5 – 1.8 |
| 2003 QW$_{111}$ (1) | 0.42 – 0.63 | 45 – 53 | 120,000 | 230,000 | 0.4 – 1.5 |
| 2003 QW$_{111}$ (2) | 0.43 – 0.62 | 134 – 128 | 120,000 | 230,000 | 0.4 – 1.5 |
| 2003 QY$_{90}$ | 0.46 – 0.84 | 32 – 58 | 32,000 | 83,000 | 0.9 – 3.2 |
| Published orbits | | | | | |
| (26308) 1998 SM$_{165}$ | 0.35 – 0.95 | 44 – 76 | 87,000 | 160,000 | 0.5 – 1.6 |
| 1998 WW$_{31}$ (1) | 0.31 – 0.90 | 37 – 69 | 23,000 | 47,000 | 1.6 – 1.9 |
| 1998 WW$_{31}$ (2) | 0.31 – 0.90 | 144 – 112 | 23,000 | 47,000 | 1.6 – 1.9 |
| 1999 OJ$_4$ (1) | 0.35 – 0.73 | 44 – 58 | 110,000 | 190,000 | 0.5 – 0.6 |
| 1999 OJ$_4$ (2) | 0.35 – 0.78 | 137 – 119 | 100,000 | 180,000 | 0.5 – 0.6 |
| 2000 QL$_{251}$ (1) | 0.46 – 0.72 | 151 – 133 | 230,000 | 530,000 | 0.3 – 1.0 |
| 2000 QL$_{251}$ (2) | 0.46 – 0.71 | 29 – 46 | 230,000 | 530,000 | 0.3 – 1.0 |
| 2001 XR$_{254}$ (1) | 0.46 – 0.56 | 21 – 29 | 73,000 | 220,000 | 0.5 – 1.7 |
| 2001 XR$_{254}$ (2) | 0.45 – 0.55 | 159 – 150 | 75,000 | 220,000 | 0.5 – 1.7 |
| 2003 TJ$_{58}$ (1) | 0.49 – 0.81 | 47 – 63 | 78,000 | 150,000 | 0.5 – 1.8 |
| 2003 TJ$_{58}$ (2) | 0.50 – 0.82 | 132 – 116 | 75,000 | 150,000 | 0.5 – 1.8 |
| 2004 PB$_{108}$ (1) | 0.07 – 0.99 | 39 – 84 | 230,000 | 440,000 | 0.4 – 1.4 |
| 2004 PB$_{108}$ (2) | 0.11 – 0.99 | 141 – 95 | 200,000 | 390,000 | 0.4 – 1.4 |

Table notes:

a. The Kozai oscillation period for $e$ and $i$ is distinct from the generally longer precession period, the time for the longitude of ascending nodes to circulate.

b. The ratio of the binary semimajor axis $a$ to the transition separation $a_c$ between oblateness-dominated dynamics and solar-tide dominated dynamics gives an indication of where Kozai oscilations are likely to be important (see Nicholson et al. 2008 section 4.1). In estimating $a_c$, we assumed a $J_2$ range of 0.001 to 0.2. Where $a/a_c > 1$, Kozai oscillations are likely to dominate the dynamics of the binary orbit. In this table, we only show results for systems where the plausible range of $a/a_c$ reached values greater than ½.

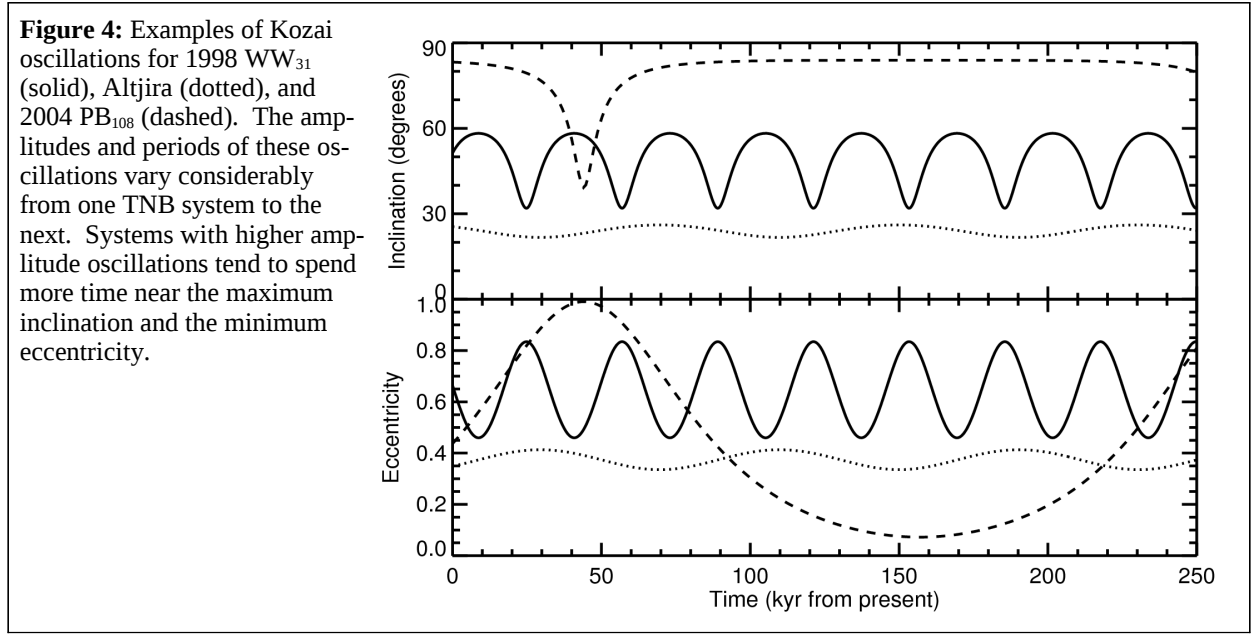

**Figure 4:** Examples of Kozai oscillations for 1998 WW$_{31}$ (solid), Altjira (dotted), and 2004 PB$_{108}$ (dashed). The amplitudes and periods of these oscillations vary considerably from one TNB system to the next. Systems with higher amplitude oscillations tend to spend more time near the maximum inclination and the minimum eccentricity.

To explore the effect of Kozai oscillations on the eccentricity distribution discussed earlier, we evaluated the eccentricity distribution over time for 250 kyr, long enough to average over the oscillations of the systems where Kozai effects are likely to be important. The resulting mean counts and their standard deviations are shown as stars with error bars in Fig. 2. Over this time scale, the eccentricity distribution remains much as we see it today. However, several systems do reach extremely high eccentricites, such as 2004 PB$_{108}$, shown in Fig. 4. Such eccentricities seem to be incompatible with the long term survival of these systems as binaries. We have already noted that higher order terms in the bodies' gravitational fields could interfere with the Kozai mechanism. We also neglected tidal dissipation, which could become important during phases of higher eccentricity (e.g., Perets and Naoz 2009, and see section 4.2 above). Accounting for these additional factors, these systems might not actually evolve through periods of such high eccentricity, remaining instead more like we observe them today. Work is in progress to investigate this possibility.

**4.4 Angular Momentum and Binding Tightness**

The specific angular momentum $J/J'$ of a binary system can provide a clue to its origin. Binaries produced via impact disruption should have $J/J' < 0.8$ (e.g., Canup 2005; Chiang et al. 2007; Descamps and Marchis 2008). As before, we assume plausible densities fall into the range 0.5 to 2.0 g cm$^{-3}$, resulting in a plausible range of $J'$ values. Also, we do not know the spin component of the total angular momentum, so we report $J_{orb}/J'$. For widely separated, near-equal binaries, $J_{spin}$ should be small compared with $J_{orb}$, even if both component spins were aligned with the orbit and near the breakup rate, but for more unequal binaries, $J_{spin}$ could exceed $J_{orb}$. For example, assuming 2 hour spins and maximum radii, for 2001 QY$_{297}$, $J_{spin}$ could only contribute up to a fifth of the total angular momentum while for Orcus-Vanth it could contribute as much as 90%. Despite these uncertainties, three systems stand out for having very low values of the specific orbital angular momentum: Eris, Orcus, and Salacia. These are also the three most massive systems in our sample, with $M_{sys} > 10^{20}$ kg and $R_1 > 300$ km. They seem likely to have formed via impacts. Most of the other systems have $J_{orb}/J' \geq 1$, favoring formation through capture, rather than impact disruption. 26309 stands out as a possible lower-mass collisional product, with the important caveat that the validity of the 0.8 threshold has yet to be established for impact speeds much greater than escape velocity. The lower mass objects in our sample would have low escape velocities, possibly much lower than typical impact velocities.

The binding tightness of a binary can be characterized in terms of the mean separation relative to the Hill radius $a/r_H$. Systems with separations approaching the Hill radius are easily disrupted, while systems with much smaller values of $a/r_H$ are comparatively robust (e.g., Petit and Mousis 2004). Plotting $a/r_H$ versus the excitation of the heliocentric orbits as measured by $\sqrt{(\sin(i_\odot)^2 + e_\odot^2)}$ reveals an intriguing pattern, as shown in Fig. 5. The loosest binaries are all on relatively quiescent heliocentric orbits, while the binaries with the most excited heliocentric orbits are all relatively tightly bound. A large area of the plot with widely separated binaries on excited heliocentric orbits is unpopulated, despite widely-separated binaries being the easiest binaries to discover. Numerous TNOs on excited heliocentric orbits have been imaged by HST, revealing no loose binaries (e.g., Noll et al. 2008b), so this does not seem to be an effect of observational bias, at least not for the size and mass range of TNBs surveyed to date. If there were no relationship whatsoever between excitation and $a/r_H$, we could scramble the pairing of excitation and $a/r_H$ values without changing the general character of the plot. However, 97.6% of the time, such a scrambling places one or more points into the empty region, implying a slightly greater than 2-σ confidence of a non-random relation between excitation and $a/r_H$. Another way of assessing statistical significance is to split the sample into two groups and compare them using a two-tailed Kolmogorov-Smirnov test. Comparing the Classical TNBs (colored red in the plot) with all other TNBs, this test says they are drawn from distinct parent populations in terms of their binding tightness at a comparable 96.4% confidence level.

What could cause a paucity of loose binaries among the more excited populations? It could be a result of processes responsible for exciting the heliocentric orbits of dynamically hot TNOs having also acted to disrupt loose binaries (e.g., Parker and Kavelaars 2010). Or it could arise from different primordial binary formation environments between objects that were to be subsequently excited and those which formed in regions that would never become excited. More data are clearly needed both to validate this observation and to shed light on its cause. Three more very loose TNB systems, 2000 $CF_{105}$ (Noll et al. 2002), 2001 $QW_{322}$ (Petit et al. 2008), and 2005 $EO_{304}$ (Kern 2005) can eventually be added to this plot. They are not included at present because their mutual orbit periods, semimajor axes, and masses have yet to be determined (according to the criteria in Section 2), but we already know that they will appear on the upper left side of this diagram, since they all have wide separations and low masses coupled with low heliocentric excitations of 0.050, 0.064, and 0.070 respectively. Their eventual inclusion will boost confidence in the pattern reported here.

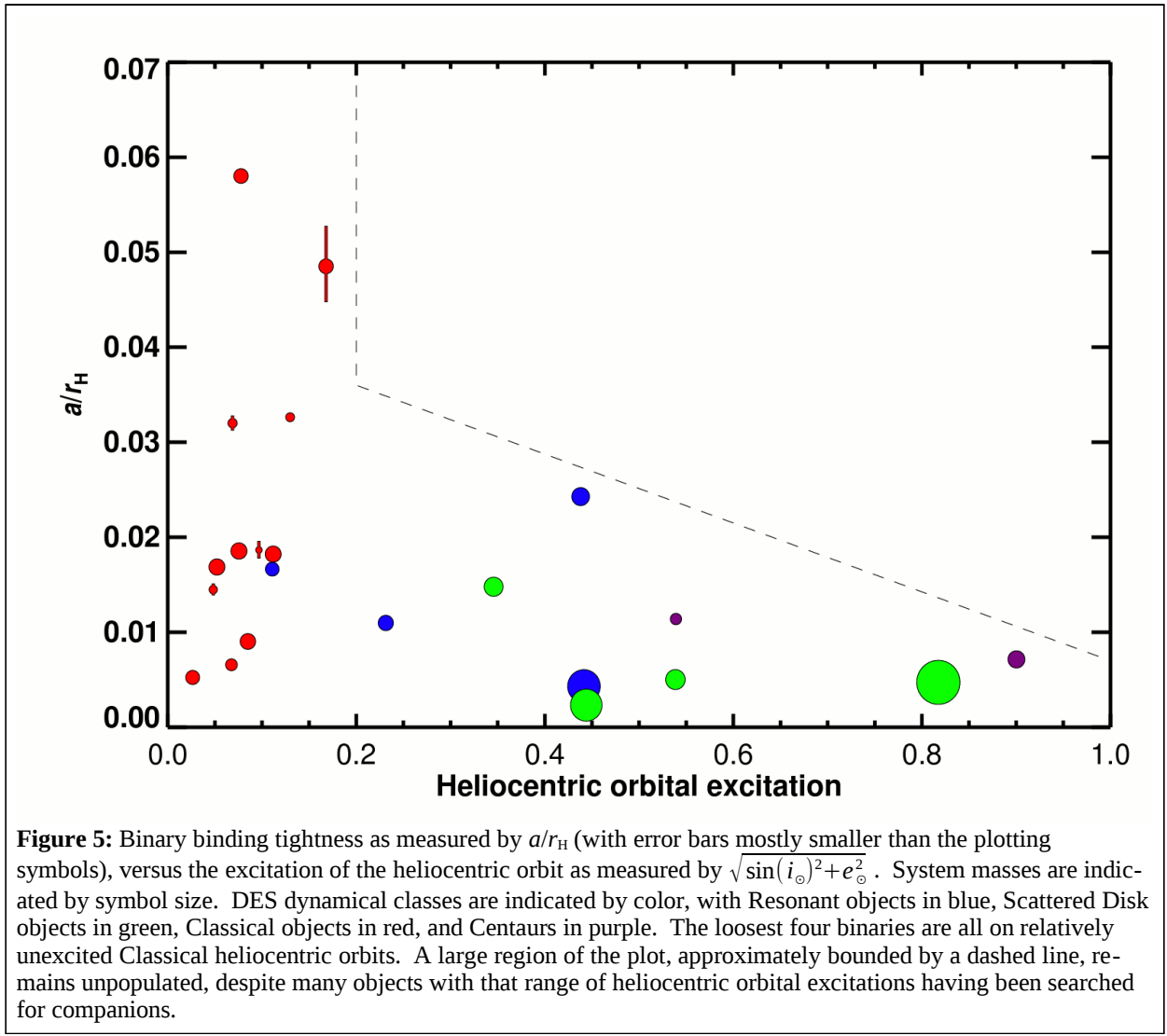

**Figure 5:** Binary binding tightness as measured by $a/r_H$ (with error bars mostly smaller than the plotting symbols), versus the excitation of the heliocentric orbit as measured by $\sqrt{\sin(i_\odot)^2 + e_\odot^2}$. System masses are indicated by symbol size. DES dynamical classes are indicated by color, with Resonant objects in blue, Scattered Disk objects in green, Classical objects in red, and Centaurs in purple. The loosest four binaries are all on relatively unexcited Classical heliocentric orbits. A large region of the plot, approximately bounded by a dashed line, remains unpopulated, despite many objects with that range of heliocentric orbital excitations having been searched for companions.

## 5. Conclusion

We report new, spatially resolved observations of eight transneptunian binary systems using HST as well as the Keck II telescope with laser guide star adaptive optics. These observations enable us to compute five new mutual orbits and improve the mutual orbits for three more, bringing the total to 22 transneptunian binary systems for which at least the mutual period, semi-major axis, and eccentricity are known. This growing sample of orbits reveals some intriguing statistical patterns likely to shed light on the formation circumstances and subsequent dynamical history or binaries in the Kuiper belt. No evidence of preferentially retrograde mutual orbits is seen, as would be expected if initially loosely-bound captures were gradually tightened through dissipation from a sea of small particles ($L^2$s mechanism). The observed distribution of orbital orientations is consistent with a random or preferentially prograde distribution. One consequence of such a random-looking distribution is that many transneptunian binary mutual orbits have very high inclinations with respect to their heliocentric orbits. For looser systems, these orientations could result in large amplitude Kozai oscillations in their mutual inclinations and eccentricities. The very existence of such high inclination systems suggests that other factors mediate the Kozai effect to prevent mutual eccentricities from reaching the extreme levels that n-body integrations predict.

## Acknowledgments

This work is based in part on NASA/ESA Hubble Space Telescope Cycle 16 program 11178 observations, with additional data from programs 9060, 9386, 9585, 9746, 9991, 10508, 10514, and 10800 obtained from the Data Archive at the Space Telescope Science Institute (STScI). STScI is operated by the Association of Universities for Research in Astronomy, Inc., under NASA contract NAS 5-26555. Support for program 11178 was provided by NASA through a grant from STScI. We are especially grateful to Tony Roman as well as Beth Perriello and Alison Vick at STScI for their quick action in scheduling HST follow-up observations. Additional data were obtained at the W.M. Keck Observatory, which is operated as a scientific partnership among the California Institute of Technology, the University of California, and NASA and made possible by the generous financial support of the W.M. Keck Foundation. The authors wish to recognize and acknowledge the very significant cultural role and reverence of the summit of Mauna Kea within the indigenous Hawaiian community. We are most fortunate to have the opportunity to conduct observations from this mountain. We especially thank Support Astronomers Hien Tran and Mark Kassis and Observing Assistants Carolyn Parker, Heather Hershley, and Cynthia Wilburn for their help with the Keck 2 telescope and Al Conrad for his timely implementation of the new differential tracking mode which was crucial to the success of our observations. We are grateful to have benefited from thorough and constructive reviews by Jean-Marc Petit and an anonymous reviewer. Finally, we thank the free and open source software communities for empowering us with key tools used to complete this project, notably Linux, the GNU tools, LibreOffice (formerly OpenOffice), MySQL, Evolution, Python, and FVWM.

## REFERENCES

Astakhov, S.A., E.A. Lee, and D. Farrelly 2005. Formation of Kuiper-belt binaries through multiple chaotic scattering encounters with low-mass intruders. *Mon. Not. R. Astron. Soc.* **360,** 401-415.

Benecchi, S.D., K.S. Noll, W.M. Grundy, M.W. Buie, D. Stephens, and H.F. Levison 2009. The correlated colors of transneptunian binaries. *Icarus* **200,** 292-303.

Brown, M.E., and E.L. Schaller 2007. The mass of dwarf planet Eris. *Science* **316,** 1585-1585.

Brown, M.E., D. Ragozzine, J. Stansberry, and W.C. Fraser 2010. The size, density, and formation of the Orcus-Vanth system in the Kuiper belt. *Astron. J.* **139,** 2700-2705.

Brucker, M.J., W.M. Grundy, J.A. Stansberry, J.R. Spencer, S.S. Sheppard, E.I. Chiang, and M.W. Buie 2009. High albedos of low inclination Classical Kuiper belt objects. *Icarus* **201,** 284-294.

Canup, R.M. 2005. A giant impact origin of Pluto-Charon. *Science* **307,** 546-550.

Chambers, J.E. 1999. A hybrid symplectic integrator that permits close encounters between massive bodies. *Mon. Not. R. Astron. Soc.* **304,** 793-799.

Chiang, E., Y. Lithwick, R. Murray-Clay, M. Buie, W. Grundy, and M. Holman 2007. A brief history of transneptunian space. In *Protostars and Planets V*. Eds. B. Reipurth, D. Jewitt, and K. Keil. Univ. of Arizona Press, Tucson, pp. 895-911.

Descamps, P., and F. Marchis 2008. Angular momentum of binary asteroids: Implications for their possible origin. *Icarus* **193,** 74-84.

Elliot, J.L., S.D. Kern, K.B. Clancy, A.A.S. Gulbis, R.L. Millis, M.W. Buie, L.H. Wasserman, E.I. Chiang, A.B. Jordan, D.E. Trilling, and K.J. Meech 2005. The Deep Ecliptic Survey: A search for Kuiper belt objects and Centaurs. II. Dynamical classification, the Kuiper belt plane, and the core population. *Astron. J.* **129,** 1117-1162.

Fabrycky, D., and S. Tremaine 2007. Shrinking binary and planetary orbits by Kozai cycles with tidal friction. *Astrophys. J.* **669,** 1298-1315.

Farrelly, D., D. Hestroffer, S.A. Astakhov, E.A. Lee, J. Berthier, F. Vachier, F. Merlin, A. Doressoundiram, and F. Marchis 2006. Orbital and physical characterization of Centaur and TNO binaries. *Bull. Amer. Astron. Soc.* **38,** 1300 (abstract).

Funato, Y., J. Makino, P. Hut, E. Kokubo, and D. Kinoshita 2004. The formation of Kuiper-belt binaries through exchange reactions. *Nature* **427,** 518-520.

Ghez, A.M., S. Salim, N.N. Weinberg, J.R. Lu, T. Do, J.K. Dunn, K. Matthews, M.R. Morris, S. Yelda, E.E. Becklin, T. Kremenek, M. Milosavljevic, and J. Naiman 2008. Measuring distance and properties of the Milky Way's central supermassive black hole with stellar orbits. *Astrophys. J.* **689,** 1044-1062.

Gladman, B., B.G. Marsden, and C. VanLaerhoven 2008. Nomenclature in the outer Solar System. In *The Solar System Beyond Neptune*, Eds. A. Barucci, H. Boehnhardt, D. Cruikshank, and A. Morbidelli. Univ. of Arizona Press, Tucson AZ, pp. 43-57.

Goldreich, P., and S. Soter 1966. Q in the solar system. *Icarus* **5,** 375-389.

Goldreich, P., Y. Lithwick, and R. Sari 2002. Formation of Kuiper-belt binaries by dynamical friction and three-body encounters. *Nature* **420,** 643-646.

Grundy, W.M., K.S. Noll, and D.C. Stephens 2005. Diverse albedos of small transneptunian objects. *Icarus* **176,** 184-191.

Grundy, W.M., J.A. Stansberry, K.S. Noll, D.C. Stephens, D.E. Trilling, S.D. Kern, J.R. Spencer, D.P. Cruikshank, and H.F. Levison 2007. The orbit, mass, size, albedo, and density of (65489) Ceto-Phorcys: A tidally evolved binary Centaur. *Icarus* **191,** 286-297.

Grundy, W.M., K.S. Noll, J. Virtanen, K. Muinonen, S.D. Kern, D.C. Stephens, J.A. Stansberry, and J.R. Spencer 2008. (42355) Typhon/Echidna: Scheduling observations for binary orbit determination. *Icarus* **197,** 260-268.

Grundy, W.M., K.S. Noll, M.W. Buie, S.D. Benecchi, D.C. Stephens, and H.F. Levison 2009. Mutual orbits and masses of six transneptunian binaries. *Icarus* **200,** 627-635.

Hamilton, D.P., and J.A. Burns 1992. Orbital stability zones about asteroids II: The destabilizing effects of eccentric orbits and of solar radiation. *Icarus* **96,** 43-64.

Kern, S.D. 2005. *A study of binary Kuiper belt objects*. Ph.D. thesis, Massachusetts Institute of Technology, Boston.

Kern, S.D., and J.L. Elliot 2006. Discovery and characteristics of the Kuiper belt binary 2003 QY$_{90}$. *Icarus* **183,** 179-185.

Krist, J.E., and R.N. Hook 2004. *The Tiny Tim User's Guide*. Version 6.3, Space Telescope Science Institute, Baltimore (available at `http://www.stsci.edu/software/tinytim`).

Konopacky, Q.M., A.M. Ghez, T.S. Barman, E.L. Rice, J.I. Bailey III, R.J. White, I.S. McLean, and G. Duchêne 2010. High-precision dynamical masses of very low mass binaries. *Astrophys. J.* **711,** 1087-1122.

Kozai, Y. 1962. Secular perturbations of asteroids with high inclination and eccentricity. *Astron. J.* **67,** 591-598.

Lee, E.A., S.A. Astakhov, and D. Farrelly 2007. Production of transneptunian binaries through chaos-assisted capture. *Mon. Not. R. Astron. Soc.* **379,** 229-246.

Le Mignant, D., M.A. Van Dam, A.H. Bouchez, J.C.Y. Chin, E. Chock, R.D. Campbell, A. Conrad, S. Doyle, R.W. Goodrich, E.M. Johansson, S.H. Kwok, R.E. Lafon, J.E. Lyke, C. Melcher, R.P. Mouser, D.M. Summers, P.J. Stomski Jr., C. Wilburn, and P.L. Wizinowich 2006. LGS AO at W.M. Keck Observatory: routine operations and remaining challenges. *Proc. SPIE* **6272**, 627201.

Margot, J.L., M.E. Brown, C.A. Trujillo, and R. Sari 2004. HST observations of Kuiper Belt binaries. *Bull. Amer. Astron. Soc.* **36,** 1081 (abstract).

Mohr, P.J., B.N. Taylor, and D.B. Newell 2008. CODATA recommended values of the fundamental physical constants: 2006. *Rev. Mod. Phys.* **80,** 633-730.

Murray, C.D., and S.F. Dermott 1999. *Solar System Dynamics*. Cambridge University Press, New York, 606 pp.

Naoz, S., H.B. Perets, and D. Ragozzine 2010. The observed orbital properties of binary minor planets. *Astrophys. J.* **719,** 1775-1783.

Nesvorný, D., J.L.A. Alvarellos, L. Dones, and H.F. Levison 2003. Orbital and collisional evolution of the irregular satellites. *Astron. J.* **126,** 398-429.

Nesvorný, D., A.N. Youdin, and D.C. Richardson 2010. Formation of Kuiper belt binaries by gravitational collapse. *Astron. J.* **140,** 785-793.

Nicholson, P.D., M. Cuk, S.S. Sheppard, D. Nesvorný, and T.V. Johnson 2008. Irregular satellites of the giant planets. In *The Solar System Beyond Neptune*, Eds. A. Barucci, H. Boehnhardt, D. Cruikshank, A. Morbidelli. University of Arizona Press, Tucson, pp. 411-424.

Noll, K.S., D.C. Stephens, W.M. Grundy, R.L. Millis, J. Spencer, M.W. Buie, S.C. Tegler, W. Romanishin, and D.P. Cruikshank 2002. Detection of two binary transneptunian objects, 1997 $CQ_{29}$ and 2000 $CF_{105}$, with the Hubble Space Telescope. *Astron. J.* **124,** 3424-3429.

Noll, K.S., D.C. Stephens, W.M. Grundy, D.J. Osip, and I. Griffin 2004a. The orbit and albedo of transneptunian binary (58534) 1997 $CQ_{29}$. *Astron. J.* **128,** 2547-2552.

Noll, K.S., D.C. Stephens, W.M. Grundy, and I. Griffin 2004b. The orbit, mass, and albedo of (66652) 1999 $RZ_{253}$. *Icarus* **172,** 402-407.

Noll, K.S., W.M. Grundy, E.I. Chiang, J.L. Margot, and S.D. Kern 2008a. Binaries in the Kuiper belt. In *The Solar System Beyond Neptune*, Eds. A. Barucci, H. Boehnhardt, D. Cruikshank, and A. Morbidelli. Univ. of Arizona Press, Tucson AZ, pp. 345-363.

Noll, K.S., W.M. Grundy, D.C. Stephens, H.F. Levison, and S.D. Kern 2008b. Evidence for two populations of Classical transneptunian objects: The strong inclination dependence of Classical binaries. *Icarus* **194,** 758-768.

Osip, D.J., S.D. Kern, and J.L. Elliot 2003. Physical characterization of the binary Edgeworth-Kuiper belt object 2001 $QT_{297}$. *Earth, Moon, & Planets* **92,** 409-421.

Parker, A., and J.J. Kavelaars 2010. Destruction of binary minor planets during Neptune scattering. *As-*

*trophys. J. Lett.* **722,** L204-L208.

Perets, H.B., and S. Naoz 2009. Kozai cycles, tidal friction, and the dynamical evolution of binary minor planets. *Astrophys. J.* **699,** L17-L21.

Petit, J.M., and O. Mousis 2004. KBO binaries: How numerous are they? *Icarus* **168,** 409-419.

Petit, J.M., J.J. Kavelaars, B.J. Gladman, J.L. Margot, P.D. Nicholson, R.L. Jones, J.W. Parker, M.L.N. Ashby, A. Campo Bagatin, P. Benavidez, J. Coffey, P. Rousselot, O. Mousis, and P.A. Taylor 2008. The extreme Kuiper belt binary 2001 $QW_{322}$. *Science* **322,** 432-434.

Pravec, P., and A.W. Harris 2007. Binary asteroid population 1: Angular momentum content. *Icarus* **190,** 250-259.

Press, W.H., S.A. Teukolsky, W.T. Vetterling, and B.P. Flannery 1992. *Numerical Recipes in C*. Cambridge University Press, New York, 994 pp.

Rieke, G.H., M. Blaylock, L. Decin, C. Engelbracht, P. Ogle, E. Avrett, J. Carpenter, R.M. Cutri, L. Armus, K. Gordon, R.O. Gray, J. Hinz, K. Su, and C.N.A. Willmer 2008. Absolute physical calibration in the infrared. *Astron. J.* **135,** 2245-2263.

Schlichting, H.E., and R. Sari 2008. The ratio of retrograde to prograde orbits: A test for Kuiper belt binary formation theories. *Astrophys. J.* **686,** 741-747.

Stansberry, J.A., W.M. Grundy, S.D. Benecchi, M. Mueller, H.F. Levison, K.S. Noll, G.H. Rieke, M.W. Buie, S.B. Porter, and H.G. Roe 2010. Physical properties of trans-neptunian binaries (120347) Salacia-Actaea and (42355) Typhon-Echidna. *Icarus* (submitted).

Stephens, D.C., and K.S. Noll 2006. Detection of six transneptunian binaries with NICMOS: A high fraction of binaries in the cold classical disk. *Astron. J.* **131,** 1142-1148.

Veillet, C., J.W. Parker, I. Griffin, B. Marsden, A. Doressoundiram, M. Buie, D.J. Tholen, M. Connelley, and M.J. Holman 2002. The binary Kuiper-belt object 1998 $WW_{31}$. *Nature* **416,** 711-713.

Weidenschilling, S.J. 2002. On the origin of binary transneptunian objects. *Icarus* **160,** 212-215.